\documentclass[12pt,a4paper]{article}
\usepackage{amsmath} 
\usepackage{amssymb}
\usepackage{graphicx,epsfig}
\usepackage[mathscr]{eucal}
\usepackage{color} 
\usepackage[numbers,sort&compress]{natbib}
\usepackage{setspace}
\def\ltap{\raisebox{-.4ex}{\rlap{$\,\sim\,$}} \raisebox{.4ex}{$\,<\,$}}

\newcommand\as{\alpha_{\mathrm{S}}}

\def\to{\rightarrow}

\def\rcut{r_{\rm cut}}
\def\mz{{m_Z}}
\def\mt{{m_{t}}}
\def\mtsquare{{m^2_{t}}}
\def\mtt{m_{t{\bar t}}}

\def\ytt{y_{t{\bar t}}}
\def\ptav{p_{T,t_{\mathrm{av}}}}
\def\yav{y_{t_{\mathrm{av}}}}
\def\yth{y_{t_\text{had}}}
\def\pth{p_{T,t_\text{had}}}
\def\ptmin{p_{T,{\rm min}}}
\def\mtmin{m_{T,{\rm min}}}
\def\ptmax{p_{T,{\rm max}}}

\def\ptlead{p_{T,t_\text{high}}}
\def\ptsub{p_{T,t_\text{low}}}
\def\ptpair{p_{T,t{\bar t}}}
\def\mtlead{m_{T,t_\text{high}}}
\def\mtsub{m_{T,t_\text{low}}}
\def\Ht{H_{T}}
\def\mTav{m_{T,t_{\rm av}}}

\def\ttb{t {\bar t}}

\newcommand\Matrix{{\sc Matrix}}
\newcommand\Munich{{\sc Munich}}

\usepackage{hyperref}
\usepackage{footnotebackref}

\begin{document} 
\begin{titlepage}
\begin{flushright}
ZU-TH 31/19\\
\end{flushright}

\renewcommand{\thefootnote}{\fnsymbol{footnote}}
\vspace*{2cm}

\begin{center}
  {\Large \bf Top-quark pair production at the LHC: \\[1ex]
   Fully differential QCD predictions at NNLO
  }
\end{center}

\par \vspace{2mm}
\begin{center}
  {\bf Stefano Catani${}^{(a)}$, Simone Devoto${}^{(b)}$, Massimiliano Grazzini${}^{(b)}$,\\[0.2cm]
    Stefan Kallweit${}^{(c)}$}
and    
{\bf Javier Mazzitelli${}^{(b)}$}

\vspace{5mm}

${}^{(a)}$INFN, Sezione di Firenze and
Dipartimento di Fisica e Astronomia,\\[0.1cm] 
Universit\`a di Firenze,
I-50019 Sesto Fiorentino, Florence, Italy\\[0.25cm]

${}^{(b)}$Physik Institut, Universit\"at Z\"urich, CH-8057 Z\"urich, Switzerland\\[0.25cm]

$^{(c)}$Dipartimento di Fisica, Universit\`{a} degli Studi di Milano-Bicocca and\\[0.1cm] INFN, Sezione di Milano-Bicocca,
I-20126, Milan, Italy

\vspace{5mm}

\end{center}

\par \vspace{2mm}
\begin{center} {\large \bf Abstract} 

\end{center}
\begin{quote}
\pretolerance 10000

We report on a new fully differential calculation of the next-to-next-to-leading-order (NNLO) QCD
radiative corrections to the production of top-quark pairs at hadron colliders.
The calculation is performed by using the $q_T$ subtraction formalism to handle and cancel
infrared singularities in real and virtual contributions. The computation is implemented in the {\sc Matrix} framework,
thereby allowing us to efficiently compute arbitrary infrared-safe observables for stable top quarks.
We present NNLO predictions for several single- and double-differential kinematical distributions
in $pp$ collisions at the centre-of-mass energy \mbox{$\sqrt{s}=13$~TeV},
and we compare them with recent LHC data by the CMS collaboration.

\end{quote}

\vspace*{\fill}
\begin{flushleft}
June 2019
\end{flushleft}
\end{titlepage}

\section{Introduction}
\label{sec:intro}

The production of top quarks at high-energy colliders is a process of utmost importance, both in
testing the validity of the Standard Model~(SM) and in the quest for new physics.
Within the SM, the main source of top-quark events in hadronic collisions is
top-quark pair ($\ttb$) production.
The large data set delivered by the CERN LHC
enables precise measurements of the $\ttb$ production cross section as a
function of the $\ttb$ kinematics (see e.g. Refs.~\cite{Khachatryan:2015oqa,Khachatryan:2015fwh,Aad:2015hna,Aad:2015mbv,Khachatryan:2016gxp,Aaboud:2016iot,Aaboud:2016syx,Sirunyan:2017mzl,Sirunyan:2018wem,Sirunyan:2019zvx}), which can be compared with the SM predictions.
At the same time these studies have a wider relevance, since $\ttb$ production is a crucial
background in many new-physics searches.

Next-to-leading-order (NLO) QCD corrections 
for this production process were obtained thirty years ago
\cite{Nason:1987xz, Beenakker:1988bq, Beenakker:1990maa, Nason:1989zy,Mangano:1991jk}.
Beyond the on-shell approximation of $\ttb$ production, first NLO QCD studies were carried out 
within the narrow-width approximation~\cite{Melnikov:2009dn,Bernreuther:2010ny,Campbell:2012uf}.
Such NLO studies were later
performed by considering
the complete $W^+W^-b\bar b$ final states, with off-shell
leptonic~\cite{Bevilacqua:2010qb,Denner:2010jp,Denner:2012yc,Heinrich:2013qaa} and
semi-leptonic~\cite{Denner:2017kzu} decays.
In the leptonic channel the case of massive bottom quarks was investigated in
Refs.~\cite{Cascioli:2013wga,Frederix:2013gra}.
NLO QCD results for off-shell $\ttb$ production in association with an additional jet were obtained
in Refs.~\cite{Bevilacqua:2015qha,Bevilacqua:2016jfk}.
NLO electroweak~(EW) corrections for on-shell $\ttb$ production were studied in
Refs.~\cite{Kuhn:2006vh,Bernreuther:2006vg,Kuhn:2013zoa,Bernreuther:2010ny,Hollik:2011ps,Pagani:2016caq},
and a merged calculation for $\ttb+0,1\,$jets including EW corrections was presented in
Ref.~\cite{Gutschow:2018tuk}.
For the leptonic decay channel the complete NLO EW corrections to the production of the
six-particle final state are known~\cite{Denner:2016jyo}.

The calculation of the next-to-next-to-leading-order (NNLO) QCD 
corrections to the $\ttb$ total cross section was completed a few years ago
\cite{Baernreuther:2012ws, Czakon:2012zr, Czakon:2012pz, Czakon:2013goa}.
NNLO results for some differential distributions were presented in
Refs.~\cite{Czakon:2015owf,Czakon:2016ckf,Czakon:2017dip}.
This calculation was recently combined with NLO EW corrections~\cite{Czakon:2017wor}.
The $\ttb$ charge asymmetry is known at NLO \cite{Kuhn:1998kw} and NNLO \cite{Czakon:2014xsa} in QCD,
and also including NLO EW corrections \cite{Czakon:2017lgo}.
First NNLO QCD results including top-quark decays are starting to appear \cite{Behring:2019iiv}.

In the present paper we deal with on-shell $\ttb$ production in NNLO QCD.
The calculation of the $\ttb$ production cross section at this perturbative order requires 
tree-level contributions with two additional
final-state partons,
one-loop contributions with one additional parton and 
purely virtual contributions.
The required tree-level and one-loop scattering amplitudes
are known. They enter the NLO calculation of 
the associated production of a $\ttb$ pair and one jet~\cite{Dittmaier:2007wz,Dittmaier:2008uj},
but in the case of NNLO $\ttb$ production they
need to be accurately evaluated also in the infrared-singular regions where the jet becomes unresolved.
The purely virtual contributions entail the square of one-loop scattering amplitudes and the two-loop
scattering amplitudes.
The squared one-loop amplitudes are known~\cite{Korner:2008bn,Anastasiou:2008vd,Kniehl:2008fd}.
The complete computation of the two-loop amplitudes has been carried out
numerically~\cite{Czakon:2008zk,Baernreuther:2013caa}.
Partial results for these amplitudes are available in analytic
form~\cite{Bonciani:2008az,Bonciani:2009nb,Bonciani:2010mn,Bonciani:2013ywa}.
Recent progress in the computation of non-planar
two-loop master integrals~\cite{Becchetti:2019tjy,DiVita:2019lpl} indicates that
the analytic calculation can be completed in the near future.

The implementation of the various scattering amplitudes in a (fully differential)
NNLO calculation is definitely a non-trivial task because of the presence of infrared~(IR) 
divergences at intermediate stages of the calculation. 
Various methods have been proposed and used to overcome these difficulties 
at the NNLO level (the interested reader can consult the list of references in
Ref.~\cite{Bendavid:2018nar}).

Using the antenna subtraction method~\cite{GehrmannDeRidder:2005cm, Abelof:2011jv},
partial results for $\ttb$ production in the $q{\bar q}$ partonic channel
were obtained by considering the complete fermionic contributions and evaluating the
remaining contributions
in the leading-colour approximation~\cite{Abelof:2014fza, Abelof:2014jna,Abelof:2015lna}.
The complete NNLO computation of
Refs.~\cite{Baernreuther:2012ws, Czakon:2012zr, Czakon:2012pz, Czakon:2013goa,Czakon:2014xsa,Czakon:2015owf,Czakon:2016ckf,Czakon:2017dip}
was performed by using the {\sc Stripper} method~\cite{Czakon:2010td,Czakon:2011ve,Czakon:2014oma}.

In a recent paper~\cite{Catani:2019iny} we have presented a new calculation of the inclusive
$\ttb$ production cross section in NNLO QCD.
This calculation completes a previous computation
that was limited to the flavour off-diagonal partonic channels~\cite{Bonciani:2015sha}.
The calculation uses the $q_T$ subtraction formalism~\cite{Catani:2007vq} to handle and cancel
IR-singular contributions
in real and virtual corrections,
and it is now completely integrated into the {\sc Matrix} framework~\cite{Grazzini:2017mhc}.
This allows us to perform fast and efficient computations of fiducial cross sections and
(multi-)differential kinematical distributions for the production of on-shell top quarks.

In the present paper we extend the results of Ref.~\cite{Catani:2019iny} in various respects.
We present NNLO QCD predictions for several differential distributions in the transverse momenta
and rapidities of the top quarks, as well as in the invariant mass and the rapidity of the
$\ttb$ system, and we discuss the results obtained by using different scale choices. 
We compare these results with CMS measurements in the lepton+jets channel at the centre-of-mass energy
\mbox{$\sqrt{s}=13$~TeV}~\cite{Sirunyan:2018wem}.
We then consider double-differential distributions and compare our results with the corresponding
measurements by CMS~\cite{Sirunyan:2018wem}.

The paper is organized as follows. In Section~\ref{sec:matrix} we illustrate the framework
in which the calculation is performed. In Section~\ref{sec:resu}
we present results for single-differential and double-differential distributions,
and we compare them with the experimental measurements.
Finally, in Section~\ref{sec:summary} we present our conclusions.
In Appendix~\ref{sec:validation} we present a quantitative comparison of our NNLO differential results with those available
in the literature.

\section[Calculation within the {\sc Matrix} framework]{Calculation within the M{\normalsize ATRIX} framework}
\label{sec:matrix}

Our fully differential NNLO computation of $\ttb$ production is carried out within
the \Matrix{}~\cite{Grazzini:2017mhc} framework.
\Matrix{} features a completely automated implementation of the $q_T$ subtraction formalism~\cite{Catani:2007vq}
to compute NNLO corrections,
and it is thus applicable to the production of an arbitrary set of colourless final-state particles
in hadronic collisions~\cite{Catani:2013tia},
as long as the two-loop virtual corrections to the corresponding leading-order~(LO) process are provided.
With appropriate modifications of the NNLO subtraction counterterm
and the explicit computation of additional soft contributions (see below),
\Matrix{} can now also deal with the production of heavy-quark pairs.

According to the $q_T$ subtraction method, the NNLO
differential cross section $d{\sigma}^{t{\bar t}}_{\rm NNLO}$ for 
the production process $pp\to t{\bar t}+X$ can be written as
\begin{equation}
\label{eq:main}
d{\sigma}^{t{\bar t}}_{\rm NNLO}={\cal H}^{t{\bar t}}_{\rm NNLO}\otimes d{\sigma}^{t{\bar t}}_{\rm LO}
+\left[ d{\sigma}^{t{\bar t}+\rm{jet}}_{\rm NLO}-
d{\sigma}^{t{\bar t}, \, CT}_{\rm NNLO}\right],
\end{equation}
where $d{\sigma}^{t{\bar t}+\rm{jet}}_{\rm NLO}$ is the $\ttb$+jet cross section at NLO accuracy.

The square bracket term of Eq.~(\ref{eq:main}) is IR finite in the limit
in which the transverse momentum of the $\ttb$ pair, $q_T$, vanishes.
However, the individual contributions
$d{\sigma}^{t{\bar t}+\rm{jet}}_{\rm NLO}$ and
$d{\sigma}^{t{\bar t}, \, CT}_{\rm NNLO}$ are separately divergent.
The contribution $d{\sigma}^{t{\bar t}+\rm{jet}}_{\rm NLO}$ can be evaluated with any available NLO method to handle and cancel IR divergences.
The IR subtraction counterterm $d{\sigma}^{t{\bar t}, \,CT}_{\rm NNLO}$
is obtained from the NNLO perturbative expansion 
(see e.g.\ Refs.~\cite{Bozzi:2005wk,Bozzi:2007pn,Bonciani:2015sha})
of the resummation formula
of the logarithmically-enhanced
contributions to the $q_T$ distribution
of the $\ttb$ pair~\cite{Zhu:2012ts,Li:2013mia,Catani:2014qha}:
the explicit form of $d{\sigma}^{t{\bar t}, \,CT}_{\rm NNLO}$ is fully known.

To complete the NNLO calculation, the second-order functions
${\cal H}^{t{\bar t}}_{\rm NNLO}$ in Eq.~(\ref{eq:main}) are needed.
These functions embody
process-independent and process-dependent contributions. The process-independent contributions to 
${\cal H}^{t{\bar t}}_{\rm NNLO}$ are analogous to those entering Higgs
boson~\cite{Catani:2007vq} and vector-boson~\cite{Catani:2009sm} production,
and they are explicitly known~\cite{Catani:2011kr,Catani:2012qa,Catani:2013tia,Gehrmann:2012ze,Gehrmann:2014yya}.
Since in $\ttb$ production both the $gg$ and the $q{\bar q}$ partonic channels contribute
at the same perturbative order,
all these process-independent contributions are required.
In the flavour off-diagonal channels 
the process-dependent contributions to ${\cal H}^{t{\bar t}}_{\rm NNLO}$
involve only amplitudes of the partonic
processes \mbox{$q{\bar q} \to t{\bar t}$} and \mbox{$gg\to t{\bar t}$} up to the one-loop level,
and the explicit results on the NLO {\em azimuthal-correlation} terms
in the transverse-momentum resummation formalism~\cite{Catani:2014qha}.
The computation of ${\cal H}^{t{\bar t}}_{\rm NNLO}$ in the flavour diagonal $q{\bar q}$ and $gg$
channels additionally requires the two-loop amplitudes
for \mbox{$q{\bar q}\to t{\bar t}$} and \mbox{$gg\to t{\bar t}$},
and the evaluation of new contributions of purely {\it soft} origin.
The two-loop amplitudes are available in a numerical form~\cite{Baernreuther:2013caa},
and the corresponding grids have been implemented into {\sc Matrix} through a suitable
interpolation routine.
The computation of the additional soft contributions
has been completed by some of us~\cite{inprep}\footnote{An independent computation of
these soft contributions is presented in Ref.~\cite{Angeles-Martinez:2018mqh}.},
and it has been implemented into {\sc Matrix} as well.

The core of the \Matrix{} framework is the Monte Carlo program \Munich{}\footnote{\Munich{} is the 
abbreviation of ``MUlti-chaNnel Integrator at Swiss~(CH) precision'' --- an automated parton-level
NLO generator by S.~Kallweit.}, which includes a fully automated implementation of the
dipole-subtraction method for massless~\cite{Catani:1996jh,Catani:1996vz}
and massive~\cite{Catani:2002hc} partons,
and an efficient phase-space integration.
All the required (spin- and colour-correlated) tree-level and one-loop (squared) amplitudes
are obtained by using {\sc OpenLoops~2}~\cite{Cascioli:2011va,openloops2},
except for the four-parton tree-level colour correlations that are based on an analytic implementation.
{\sc OpenLoops~2} relies on its new on-the-fly tensor reduction~\cite{Buccioni:2017yxi} that guarantees
stability all over the phase space, especially in the IR-singular regions,
while scalar integrals from {\sc Collier}~\cite{Denner:2014gla,Denner:2016kdg} are used.
To the purpose of validating our results for the real--virtual corrections,
we have also used the independent matrix-element generator {\sc Recola}~\cite{Actis:2016mpe,Denner:2017wsf}, which employs tensor reduction and scalar integrals from {\sc Collier}, and we find complete agreement.

The subtraction in the square brackets of Eq.~(\ref{eq:main}) is not local, but the cross section 
is formally finite in the limit \mbox{$q_T \to 0$}. In practice, a technical cut on $q_T$
is introduced to render $d{\sigma}^{\ttb+\mathrm{jet}}_{\mathrm{(N)LO}}$ and 
$d{\sigma}^{\mathrm{CT}}_{\mathrm{(N)NLO}}$ separately finite.
Therefore, in our actual implementation, the $q_T$ subtraction method is
very similar to a phase-space slicing method. 
It turns out that a cut, $r_{\mathrm{cut}}$, on the 
dimensionless quantity \mbox{$r=q_T/\mtt$} ($\mtt$
denotes the invariant mass of the $\ttb$ pair) 
is more convenient from a practical point of view. 
The absence of any residual logarithmic
dependence on $r_{\mathrm{cut}}$ is a strong evidence of the correctness of the 
computation, since any mismatch between the contributions would result in a divergence 
of the cross section in the limit \mbox{$r_{\mathrm{cut}}\to0$}.
The remaining power-suppressed contributions vanish in that limit, and they can be controlled by
monitoring the $r_{\mathrm{cut}}$ dependence of the cross section.

The $r_\text{cut}\to 0$ extrapolation for the total cross section is carried out by using the approach
introduced in Ref.~\cite{Grazzini:2017mhc}.
A quadratic least $\chi^2$ fit to the $r_\text{cut}$ dependent results is performed and repeated
by varying the upper bound of the $r_\text{cut}$ interval.
Finally, the result with the lowest $\chi^2/$degrees-of-freedom value is taken as the best fit,
while the remaining results are used to estimate the extrapolation uncertainty.
In addition to this analysis at the level of the total cross section,
we have performed a similar bin-wise extrapolation in the computation of differential cross sections.
We find that the results are in good agreement with those obtained by directly
using a sufficiently low value of $r_\text{cut}$ (\mbox{$\rcut \lesssim 0.15\%$}).

\section{Results}
\label{sec:resu}

To present our quantitative results, we consider $pp$ collisions at \mbox{$\sqrt{s}=13$~TeV},
and we fix the pole mass $\mt$ of the top quark to the value \mbox{$\mt=173.3$~GeV}.
We consider $n_F=5$ massless quark flavours, and we use the corresponding NNPDF31~\cite{Ball:2017nwa}
sets of parton distribution functions~(PDFs) with \mbox{$\as(\mz)=0.118$}.
In particular, N$^n$LO (with $n = 0,1,2$) predictions are obtained by using PDFs
at the corresponding perturbative order and the evolution of $\as$ at \mbox{$(n + 1)$}-loop order,
as provided by the PDF set.

QCD scale uncertainties are estimated through the customary procedure of independently varying
the renormalization ($\mu_R$) and factorization ($\mu_F$) scales 
by a factor of two
around their common central value $\mu_0$ with the constraint \mbox{$0.5\leq \mu_F/\mu_R\leq 2$},
i.e.\ we use the standard 7-point scale variation.

Setting $\mu_0=\mt$, the total cross sections and their corresponding scale uncertainties read
\begin{equation}
\sigma_\text{LO}^{\ttb} = 478.9(1)^{+29.6\%}_{-21.4\%}\text{~pb}\,,\quad
\sigma_\text{NLO}^{\ttb} = 726.9(1)^{+11.7\%}_{-11.9\%}\text{~pb}\,,\quad
\sigma_\text{NNLO}^{\ttb} = 794.0(8)^{+3.5\%}_{-5.7\%}\text{~pb}\,.
\label{eq:sigmatot}
\end{equation}
We note that the LO and NLO results are not fully consistent within the corresponding uncertainties,
indicating that, at least at LO, scale variations cannot be trusted as perturbative uncertainties.
Similar features are shared by various other hard-scattering processes at hadron colliders.
In contrast, the NLO and NNLO predictions are consistent, suggesting that scale variations
can be used to estimate the size of perturbative contributions beyond NNLO.

The characteristic hard-scattering scale that controls the perturbative QCD behaviour
of the total cross section $\sigma^{\ttb}$ is $\mt$. In our calculation of
$\sigma^{\ttb}$, as reported in Eq.~(\ref{eq:sigmatot}), we have used QCD scales ($\mu_R$ and
$\mu_F$) at values of the order of $\mt$. 
Differential cross sections are 
controlled by corresponding characteristic hard scales, and we use QCD scales of that order
in our computation of these observables.
The characteristic hard scale specifically depends on the differential cross section under consideration. 

Having at our disposal a fully differential calculation we can also use {\it dynamical}
QCD scales. By dynamical we mean hard scales that refer to multi-differential
cross sections
eventually integrated over the phase space to obtain the specific 
differential cross section under consideration. 
The use of a dynamical scale produces practical simplifications
since it allows us to compute several observables (e.g., differential cross sections)
simultaneously, without changing the QCD scales on an observable-dependent basis.
In practice, we use dynamical scales that are expected to be ``effectively similar''
to characteristic hard scales. Moreover, the study of dynamical scales
is of interest independently of how we use them.

The default dynamical scale that we use throughout the paper is set to the 
central value \mbox{$\mu_0=\Ht/2$}, where $\Ht$ is the sum of the transverse masses
of the top and antitop quarks,
\begin{equation}
  \Ht=m_{T,t}+m_{T,{\bar t}}\,,
\end{equation}
with
\begin{equation}
m_{T,t({\bar t})}=\sqrt{\mtsquare+p^2_{T,t({\bar t})}}\, ,
\end{equation}
and $p_{T,t}$ and $p_{T,{\bar t}}$ are the transverse momenta of the top and the antitop quark, respectively.
We present differential cross sections that are obtained by using 
\mbox{$\mu_0=\Ht/2$} and values of $\mu_0$ of the order of the characteristic hard scale
for that cross section.  We also show results obtained by using central scales that are
lowered by a factor of $1/2$. A reduced central scale, such as $\Ht/4$, was considered
in the studies of Ref.~\cite{Czakon:2016dgf} on the basis of features of
fastest perturbative convergence of some observables,
and it was also already suggested in Ref.~\cite{Denner:2012yc}.

We have chosen the dynamical scale $\Ht$ since it is expected to be parametrically
of the same order as the characteristic hard scale of the observables that we examine
in this paper. This {\it a priori} expectation is based on the kinematical features of
these observables and on the general dynamical features of $\ttb$ production.
In the following paragraph we briefly comment about this. Independently of the expectation,
throughout the paper we comment on the actual quantitative results that we obtain by
using different QCD scales.

Owing to dynamics, the typical size of both $p_{T,t}$ and $p_{T,{\bar t}}$ is
of the order of $\mt$ (see, e.g., Figs.~\ref{fig:pt1}--\ref{fig:pth} and 
\ref{fig:pth_yth}). Therefore, in the case of observables that are inclusive over
$p_{T,t}$ and $p_{T,{\bar t}}$, such as the total cross section and the pair rapidity
distribution in Fig.~\ref{fig:ytt}, $\Ht/2$ turns out to be of the same order as 
$\mt$, which is the characteristic hard scale for these observables. Analogously,
since \mbox{$p_{T,t} \sim p_{T,{\bar t}}$}, $\Ht/2$ turns out to be of the same order
as the transverse masses, which are the characteristic hard scales for the differential cross sections
in Figs.~\ref{fig:pt1}--\ref{fig:pth} and \ref{fig:pth_yth}.
The invariant mass $\mtt$ of the $\ttb$ pair is the characteristic hard scale in the case
of the differential cross sections in Figs.~\ref{fig:mtt}, \ref{fig:ytt_mtt} and \ref{fig:mtt_pth}.
The invariant mass is of the same order as $\Ht$ with the exception of the kinematical subregions
where the transverse momentum $\ptpair$
of the pair or the rapidity separation \mbox{$|y_t - y_{\bar t}|$} between
the top and the antitop quark are large. However, these subregions are dynamically suppressed,
and therefore they give a minor contribution to the inclusive
(over $\ptpair$ and \mbox{$|y_t - y_{\bar t}|$}) cross sections
in Figs.~\ref{fig:mtt}, \ref{fig:ytt_mtt} and \ref{fig:mtt_pth}.

Our numerical results for differential cross sections are compared with the measurements of the
CMS collaboration~\cite{Sirunyan:2018wem} (the data correspond to an integrated luminosity of $35.8~{\rm fb}^{-1}$) in the lepton+jets channel at parton level.
The extrapolation from particle to parton level is carried out by the CMS collaboration in the inclusive phase space,
and therefore no kinematical cuts are applied to obtain our theoretical predictions.
To perform the comparison, our results are multiplied by the factor $0.292$,
which corresponds to the value $0.438$~\cite{Tanabashi:2018oca}
of the semileptonic decay fraction of the $\ttb$ pair, multiplied by a factor of $2/3$
since Ref.~\cite{Sirunyan:2018wem} considers only the decay into electrons and muons.

In Ref.~\cite{Sirunyan:2018wem} the CMS data for single- and double-differential distributions
are compared to theoretical results obtained with the NLO Monte Carlo event generators
{\sc POWHEG}~\cite{Nason:2004rx,Frixione:2007vw,Alioli:2010xd},
interfaced either to {\sc PYTHIA8}~\cite{Sjostrand:2007gs} or to {\sc HERWIG++}~\cite{Bahr:2008pv},
and {\sc MG5\_aMC@NLO}~\cite{Alwall:2014hca} interfaced to {\sc PYTHIA8}~\cite{Sjostrand:2007gs}
(using the {\sc FxFx} method \cite{Frederix:2012ps} to deal with multijet merging).
In addition, some of the measured parton-level single-differential distributions,
namely the transverse-momentum and rapidity distributions of the leptonically and hadronically
decaying top quark and the invariant-mass and rapidity distribution of the $\ttb$ pair,
are also compared to the NNLO QCD+NLO EW results of Ref.~\cite{Czakon:2017wor}.
None of the double-differential distributions in Ref.~\cite{Sirunyan:2018wem} are compared to theoretical results beyond NLO QCD.

\subsection{Single-differential distributions}
\label{sec:single}

In this section we present LO, NLO and NNLO results for a selection of single-differential distributions
and compare them with the CMS measurements from Ref.~\cite{Sirunyan:2018wem}.
At each perturbative order the scale-uncertainty bands in the figures
are computed as explained at the beginning of Section~\ref{sec:resu}. 

We start the presentation by considering the transverse-momentum distributions of the top and antitop quarks.
For each event we classify the transverse momenta
according to their maximum and minimum values, $\ptlead$ and $\ptsub$.

\begin{figure}[t]
\includegraphics[height=0.30\textheight]{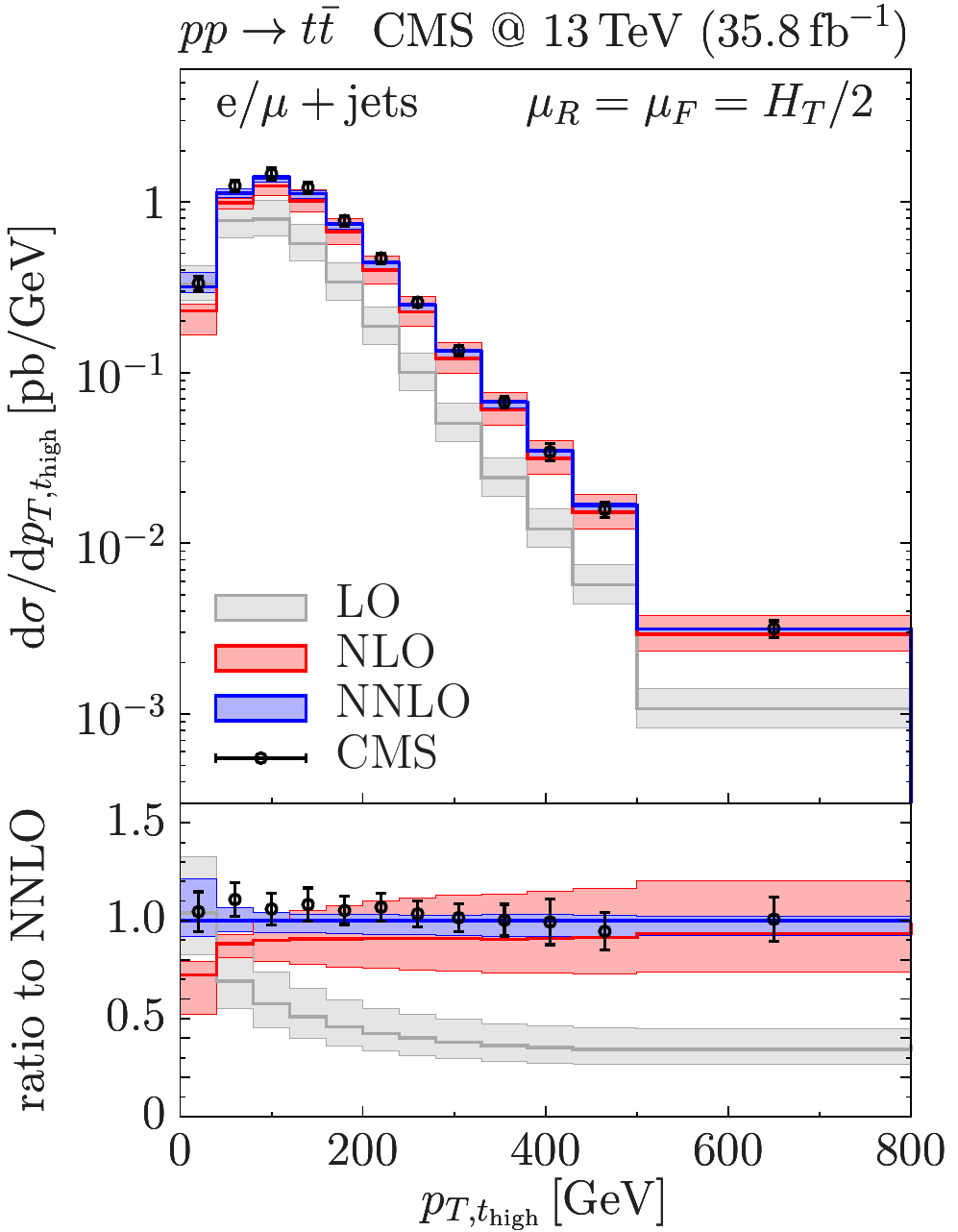}
\includegraphics[height=0.30\textheight]{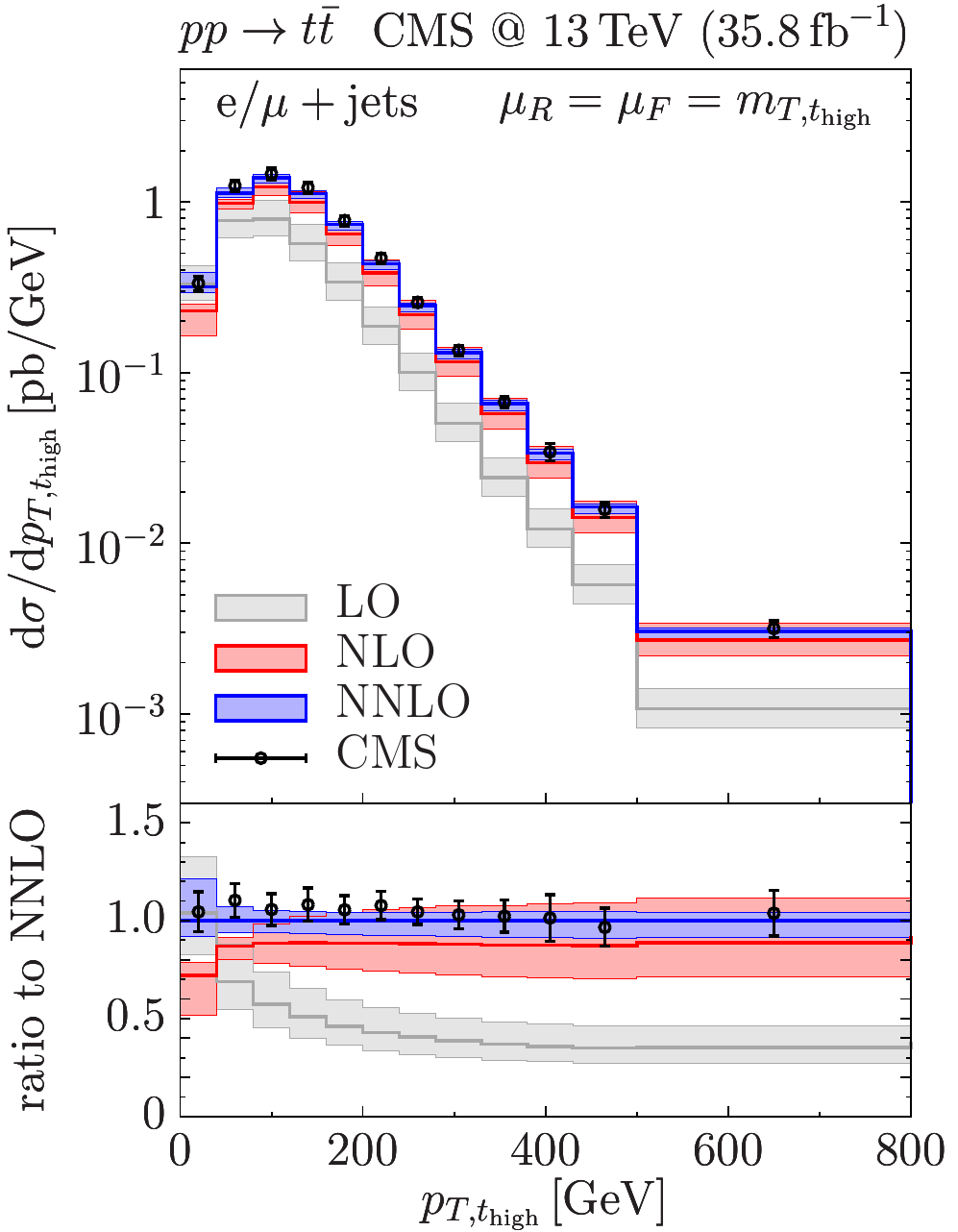}
\includegraphics[height=0.30\textheight]{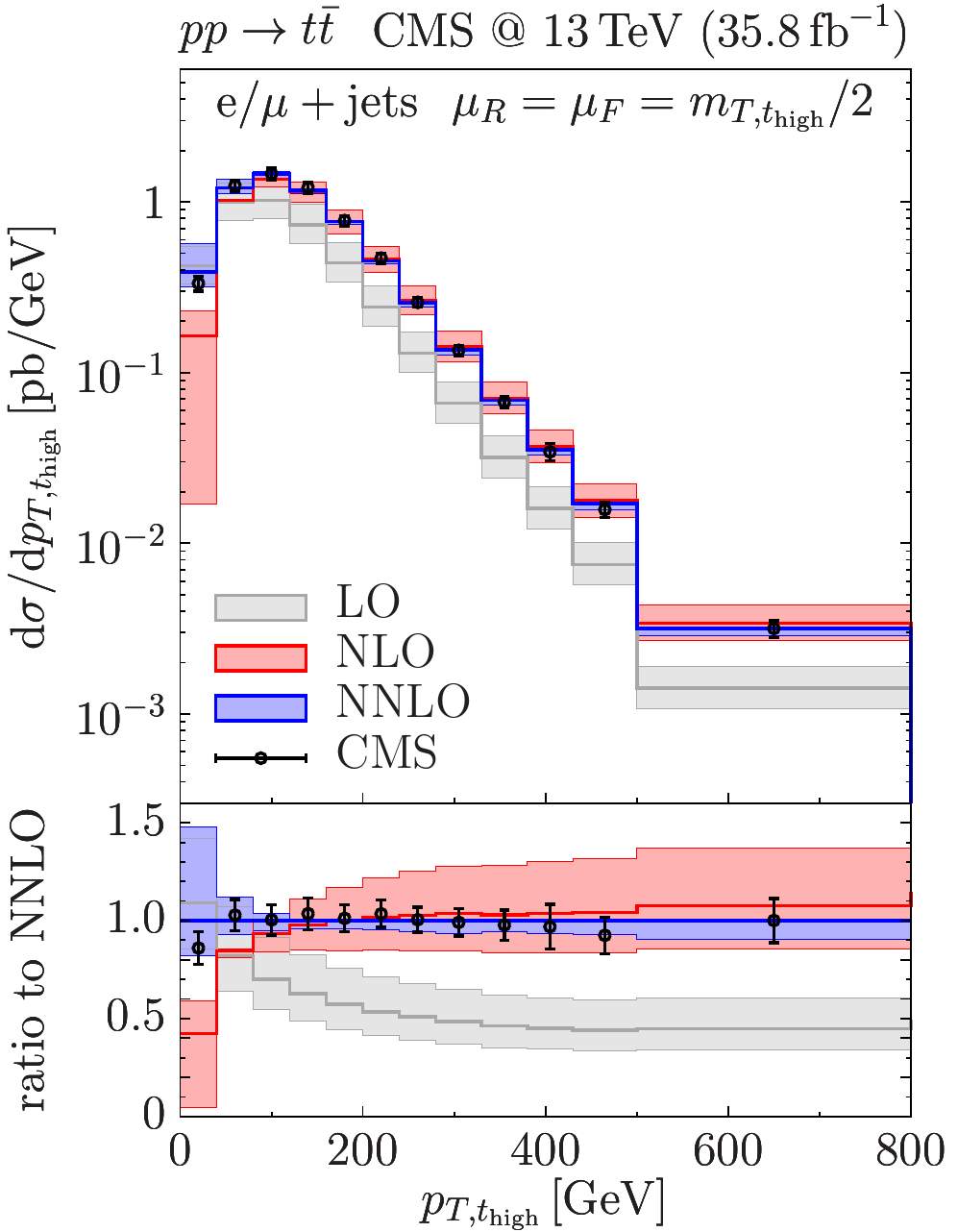}
\vspace{-4ex}
\caption{\label{fig:pt1}
  Single-differential cross sections as a function of $\ptlead$.
  CMS data~\cite{Sirunyan:2018wem} and LO, NLO and NNLO results
  for central scales equal to $\Ht/2$ (left), $\mtlead$ (central) and $\mtlead/2$ (right).
}
\end{figure}
\begin{figure}[t]
\includegraphics[height=0.30\textheight]{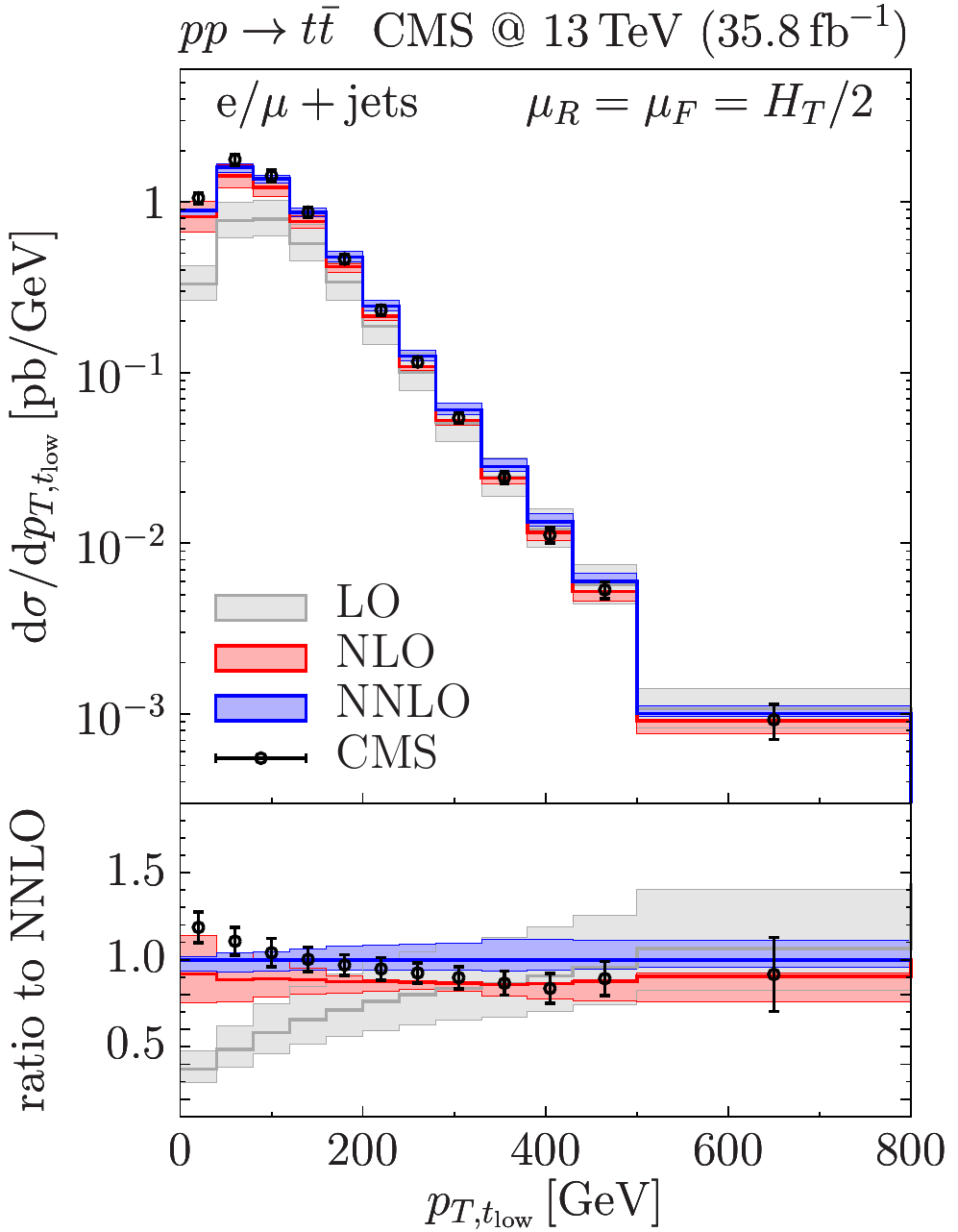}
\includegraphics[height=0.30\textheight]{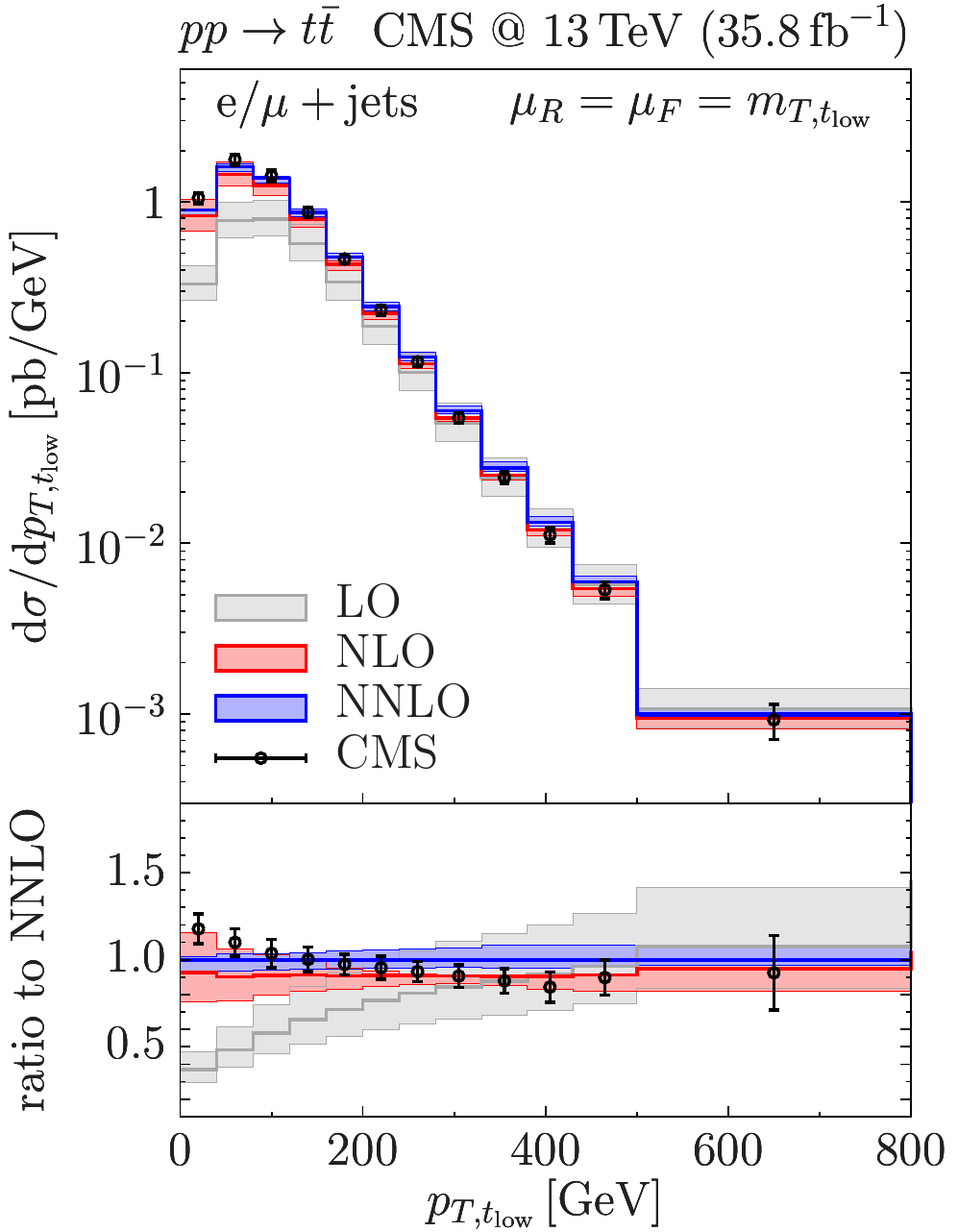}
\includegraphics[height=0.30\textheight]{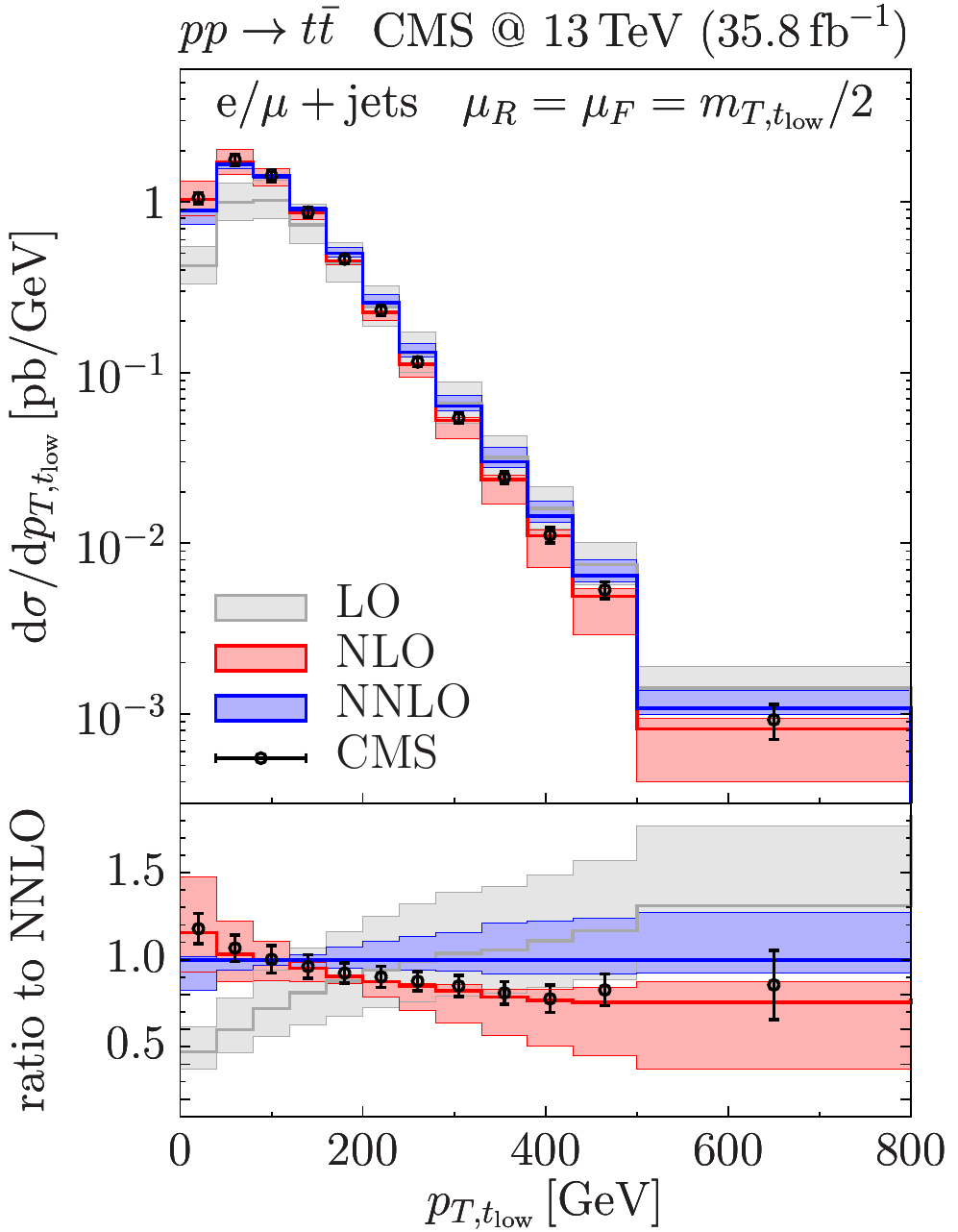}
\vspace{-4ex}
\caption{\label{fig:pt2}
  Single-differential cross sections as a function of  $\ptsub$.
  CMS data~\cite{Sirunyan:2018wem} and LO, NLO and NNLO results
  for central scales equal to $\Ht/2$ (left), $\mtsub$ (central) and $\mtsub/2$ (right).
}
\end{figure}

In Figs.~\ref{fig:pt1} and \ref{fig:pt2}~(left)
we show these distributions\footnote{NNLO results for these distributions have been recently presented in Ref.~\cite{Czakon:2019bcq}.}
computed at our reference scale \mbox{$\mu_0=H_T/2$}.
The characteristic hard scale of a transverse-momentum distribution is the transverse mass $m_T$.
Therefore, in Fig.~\ref{fig:pt1} (central and right) we also report the $\ptlead$ distribution for
central scales \mbox{$\mu_0=\mtlead$} and \mbox{$\mu_0=\mtlead/2$}, respectively,
while in Fig.~\ref{fig:pt2} (central and right) we consider $\ptsub$
for \mbox{$\mu_0=\mtsub$} and \mbox{$\mu_0=\mtsub/2$}.
The $\ptlead$ distribution is peaked at \mbox{$\ptlead\sim 100$~GeV},
while the $\ptsub$ distribution is peaked at \mbox{$\ptsub\sim 60$~GeV}.

We first discuss the $\ptlead$ distribution and focus on the \mbox{$\ptlead\to 0$} region.
If $\ptlead$ is small, both top quarks are forced to have small transverse momenta.
As a consequence, this kinematical region corresponds to a small transverse momentum of
the top-quark pair, $\ptpair$.
The small-$\ptpair$ region exhibits Sudakov-type divergences \cite{Zhu:2012ts,Li:2013mia,Catani:2014qha,Catani:2018mei} at fixed order in perturbation theory,
from the strong unbalance of real and virtual contributions due to soft-collinear emissions.
In the computation of $\ptlead$, the unphysical fixed-order behaviour of $\ptpair$ is smeared
due to the integration over $\ptsub$, and the Sudakov-type perturbative divergences disappear, by leaving (possibly large) residual effects.
The amount of smearing is controlled by the shape of the $\ptlead$ distribution in the low-$p_T$ region
at LO, which affects the unbalance between real and virtual contributions.
The steeply rising LO distribution at low $p_T$ strongly suppresses real radiation,
and the NLO radiative corrections to $\ptlead$ tend to be large and negative
as $\ptlead$ decreases (a large and positive effect occurs at NNLO, and so forth).
Accurate theoretical predictions of the detailed shape of the $\ptlead$ distribution at small $p_T$ require studies of all-order resummation effects of Sudakov type.
However, in the case of large $p_T$ bins (as is the case in Fig.~\ref{fig:pt1}), reliable predictions can be obtained by considering perturbation theory
at a sufficiently high order.

Comparing the results in Fig.~\ref{fig:pt1} for the scales $\mu_0=H_T/2$ (left) and $\mu_0=\mtlead$ (central)
we see that they are rather similar, and that the NNLO prediction agrees with the data.
The scale \mbox{$\mu_0=\mtlead/2$} also leads to good agreement with the data,
but the corresponding NNLO uncertainty band is significantly narrower,
especially in the intermediate region of transverse momenta,
which is not observed for the corresponding band at NLO.
This behaviour, namely the drastic shrinking of the scale-uncertainty band from NLO to NNLO,
might indicate that for this choice scale variations cannot be trusted
as an estimate of the perturbative uncertainties at NNLO.
We also note that in the intermediate and large $p_T$ region the result obtained by using \mbox{$\mu_0=\mtlead/2$} coincides
with the upper bound of the corresponding NNLO band  (i.e., the point $\mu_0=\mtlead/2$ corresponds to a local maximum of the NNLO cross section as a function of the scales).

We now discuss the $\ptsub$ distribution. In the region \mbox{$\ptsub\to 0$},
for LO kinematics
both the top and the antitop quark are required to have small $p_T$.
At NLO, real corrections open up a phase-space region where one top quark has a small $p_T$ and
the other one has a relatively large $p_T$,
thereby leading to large positive radiative contributions.
The perturbative instability affecting the \mbox{$\ptlead\to 0$} behaviour is now spread
over the entire region of transverse momenta since, contrary to the low-$\ptlead$ region,
small values of $\ptsub$ do not constrain $\ptpair$ to be small.
The choices \mbox{$\mu_0=H_T/2$} and \mbox{$\mu_0=\mtsub$} lead to rather similar results.
In both cases, at low and large $\ptsub$ NLO and NNLO bands overlap,
whereas they do not in the intermediate region where
the NLO band shrinks, showing that NLO perturbative uncertainties are underestimated.
We note that the scale \mbox{$\mu_0=\mtsub/2$} makes the perturbative convergence worse and the scale uncertainties
larger at both NLO and NNLO.

\begin{figure}
\includegraphics[height=0.30\textheight]{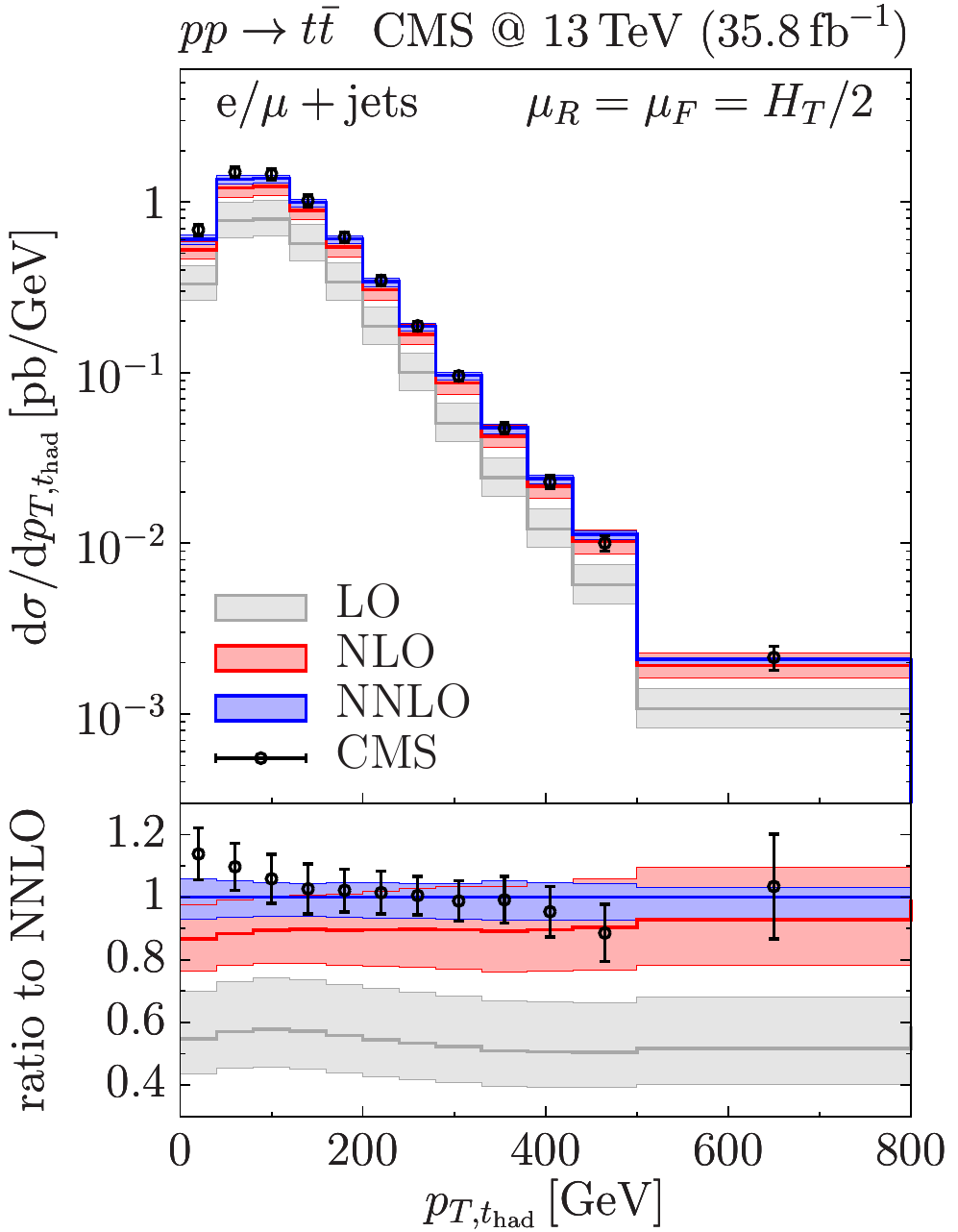}
\includegraphics[height=0.30\textheight]{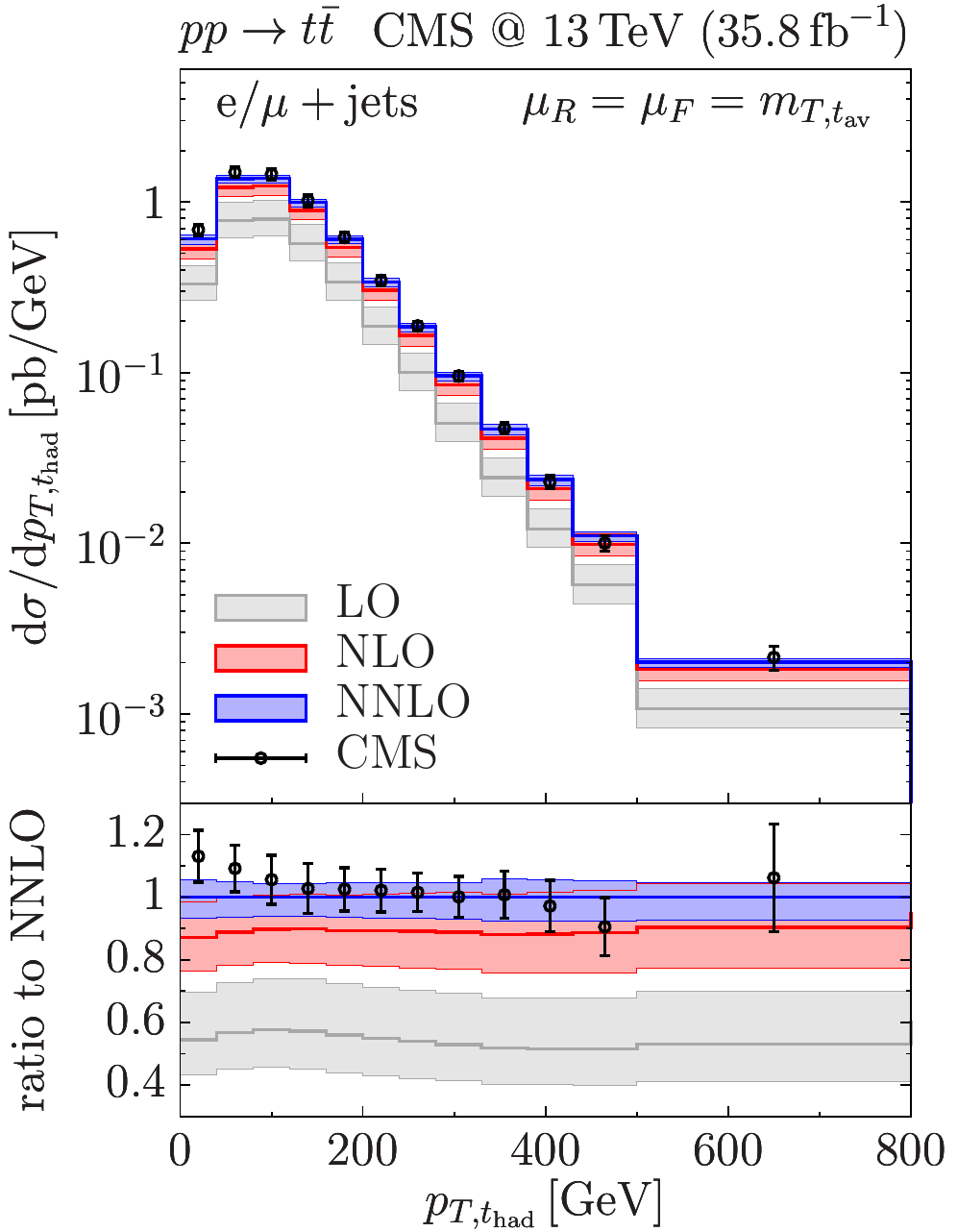}
\includegraphics[height=0.30\textheight]{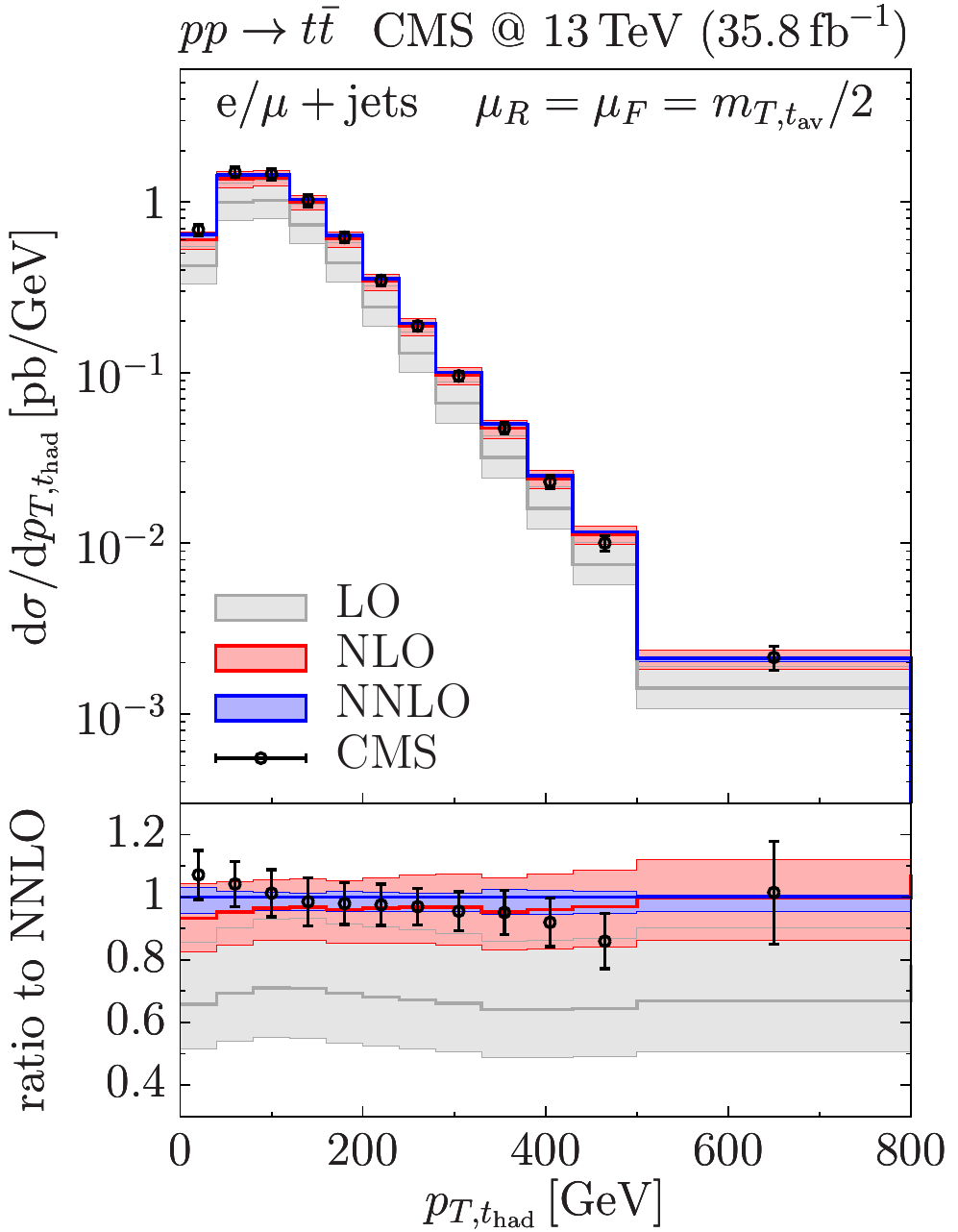}
\vspace{-4ex}
\caption{\label{fig:pth}
  Single-differential cross sections as a function of $\pth$.
  CMS data~\cite{Sirunyan:2018wem} and LO, NLO and NNLO results
  for central scales equal to $\Ht/2$ (left), $\mTav$ (central) and $\mTav/2$ (right).
}
\end{figure}

We next consider the distribution in the transverse momentum of the hadronically decaying top or antitop quark, $\pth$.
Since our calculation refers to stable top quarks, a prediction for the $\pth$ distribution
can be obtained by computing the transverse-momentum spectra of the top and the antitop quark,
and taking their average afterwards.%
\footnote{This is also the definition used in the data/theory comparison performed
by the CMS collaboration \cite{Sirunyan:2018wem}.}
Discussing our theoretical predictions, we refer to this as the $\ptav$ distribution.
The corresponding LO, NLO and NNLO results are depicted in Fig.~\ref{fig:pth}
for three different scale choices.
We show the predictions for our default choice \mbox{$\mu_0=H_T/2$} on the left.
In the predictions for our natural scale choices, 
the top (antitop) $p_T$ distributions required to compute the average are evaluated
at the corresponding transverse mass, $\mu_0=m_{T,t({\bar t})}$ (central) and $\mu_0=m_{T,t({\bar t})}/2$ (right).
We denote these scale choices as $\mTav$ and $\mTav/2$, respectively.
The $\pth$ distribution has a maximum at \mbox{$\pth \sim 80~{\rm GeV}$}.
The LO and NLO scale-uncertainty bands do not overlap, except for \mbox{$\mu_0=\mTav/2$}.
This is consistent with what happens for the corresponding total cross sections.
The NLO and NNLO bands do overlap in the entire $\pth$ range,
suggesting a good convergence of the perturbative expansion.
In Fig.~\ref{fig:pth} we also observe that the scale choices $\mu_0=H_T/2$ (left) and $\mu_0=\mTav$ (central)
give rather similar results.
On the contrary, the choice \mbox{$\mu_0=\mTav/2$} suggests a faster convergence
of the perturbative expansion \cite{Czakon:2016dgf}. 
However, we also note that with this scale choice the NLO scale dependence is similar
to what we obtain with $\mu_0=H_T/2$ and $\mu_0=\mTav$,
whereas the NNLO scale dependence is significantly smaller than at NLO,
thereby suggesting a possible underestimation of the perturbative uncertainty at NNLO.
We note that \mbox{$\mu_0=\mTav/2$} is also the scale used
for the NNLO QCD+NLO EW prediction~\cite{Czakon:2017wor}
to which the CMS data are compared in Ref.~\cite{Sirunyan:2018wem}.
The data show that the measured $\pth$ distribution is slightly softer than the NNLO prediction.
This is noticed also by the CMS collaboration~\cite{Sirunyan:2018wem} and in previous comparisons between NNLO results and LHC measurements \cite{Khachatryan:2015oqa,Khachatryan:2015fwh,Aad:2015hna,Aad:2015mbv,Khachatryan:2016gxp,Aaboud:2016iot,Aaboud:2016syx,Sirunyan:2017mzl}.
However Fig.~\ref{fig:pth} shows that the NNLO result and the data are consistent within the respective uncertainties.
Our predictions for the $p_T$ spectrum of the leptonically decaying top quark are, of course,
identical to those for $\pth$, and they are not shown here.
The comparison with the data shows similar features.

We add a few comments on the perturbative behaviour of the $\ptlead$, $\ptsub$ and $\ptav$ distributions
presented in Figs.~\ref{fig:pt1}, \ref{fig:pt2} and \ref{fig:pth}.
The three distributions are identical at LO,
but their behaviour beyond LO is clearly very different.
As we can see from Fig.~\ref{fig:pth}, the shape of the $\ptav$ distribution is almost unchanged
with respect to the LO prediction.
This feature is somehow expected, since the transverse-momentum spectrum of the top (antitop) quark
at higher orders is affected by recoiling hard multijet radiation,
which leads to a partly softer spectrum only at quite high $\ptav$
(beyond the $\pth$ range in Fig.~\ref{fig:pth}),
where hard multijet radiation is kinematically suppressed.

The physical shape of the $\ptlead$ and $\ptsub$ distributions is expected
to be different from $\ptav$ at low and intermediate values of $p_T$.
There we roughly have \mbox{$\ptlead - \ptsub \sim \ptpair$}.
The $\ptpair$ distribution, which is confined to \mbox{$\ptpair = 0$} at LO,
has an average value of about 50~GeV (which is already achieved at NLO),
and it is localized in the small-$\ptpair$ region with a peak around 10~GeV~\cite{Zhu:2012ts,Catani:2018mei}.
We thus expect that the physical shape of $\ptlead$ ($\ptsub$) is harder (softer) than its LO counterpart,
with shape distortions of few tens of GeV as given by the size of $\ptpair$.
Indeed, this is what we can observe from the comparison between the data and the LO prediction
at small and intermediate values of $p_T$ in Figs.~\ref{fig:pt1} and~\ref{fig:pt2}.
This shape distortion has a smaller effect at high values of both $\ptlead$ and $\ptsub$.

In view of these physical expectations, it is not surprising that the shape
of the $\ptlead$ and $\ptsub$ distributions is strongly affected by beyond-LO contributions.
As discussed before, their fixed-order perturbative features are a smoothened version
of the corresponding features of the $\ptpair$ distribution \cite{Catani:2018mei},
the smoother behaviour being due to the smearing that is produced by the integration
of $\ptpair$ over the respective unobserved $p_T$.

\begin{figure}
\includegraphics[height=0.30\textheight]{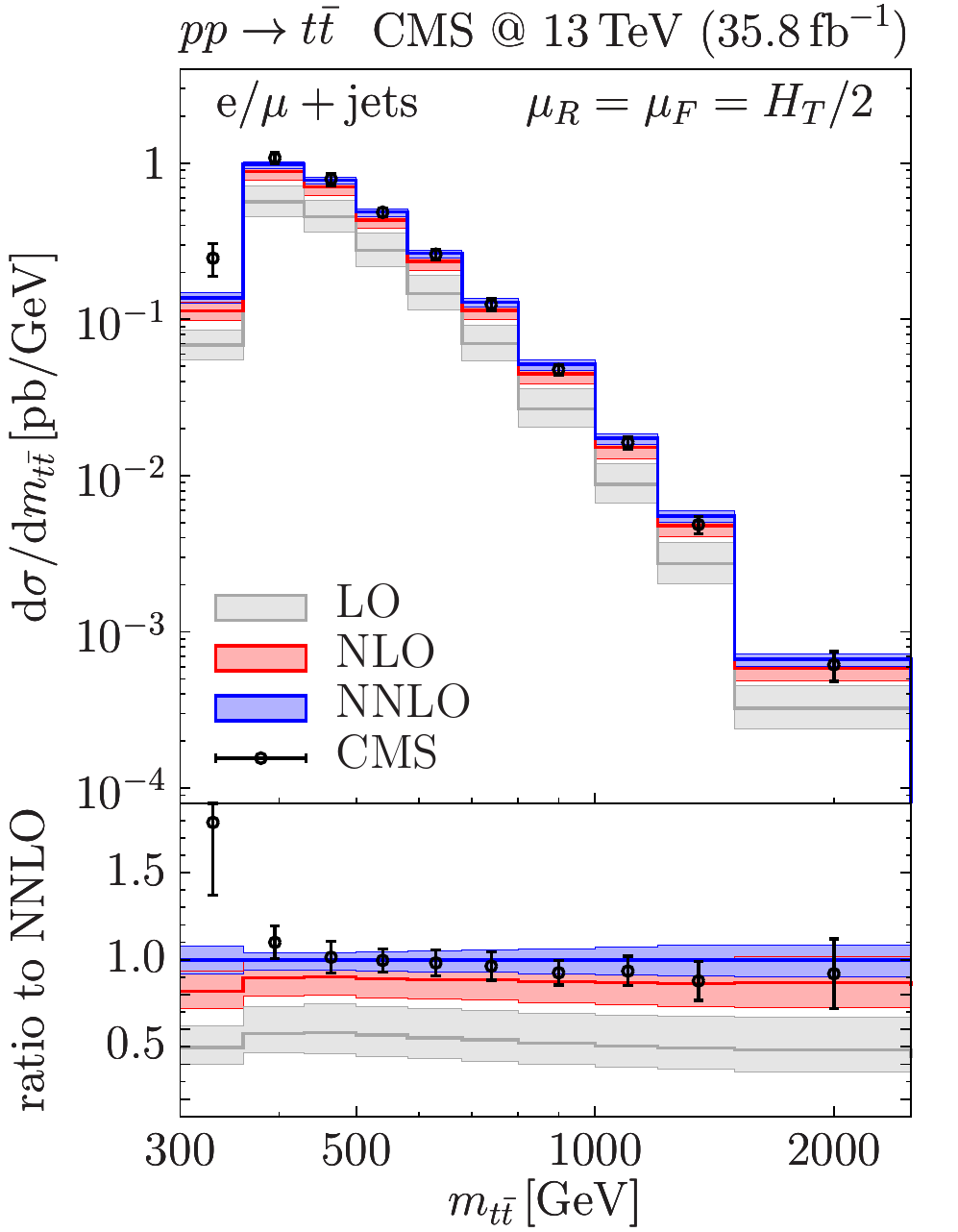}
\includegraphics[height=0.30\textheight]{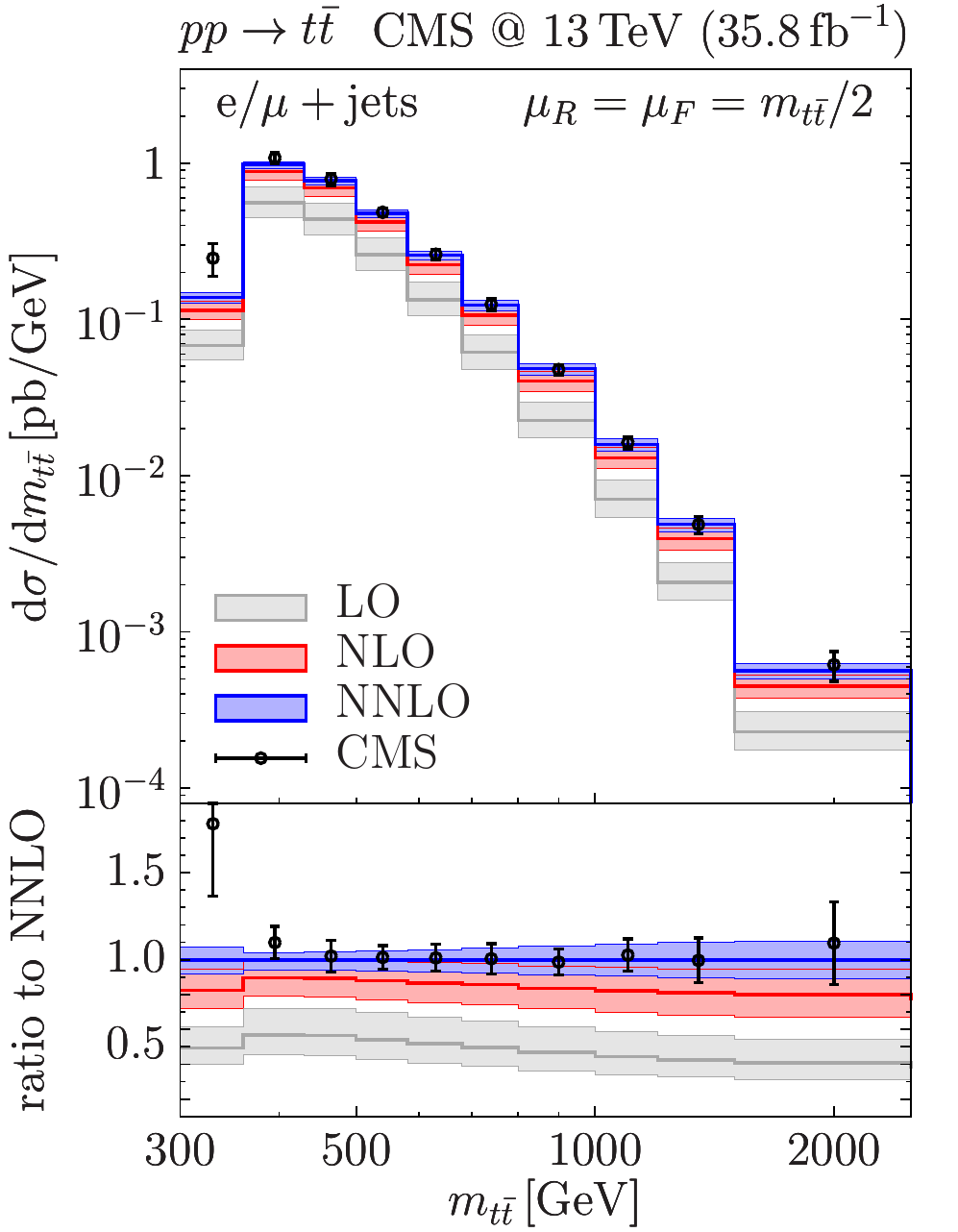}
\includegraphics[height=0.30\textheight]{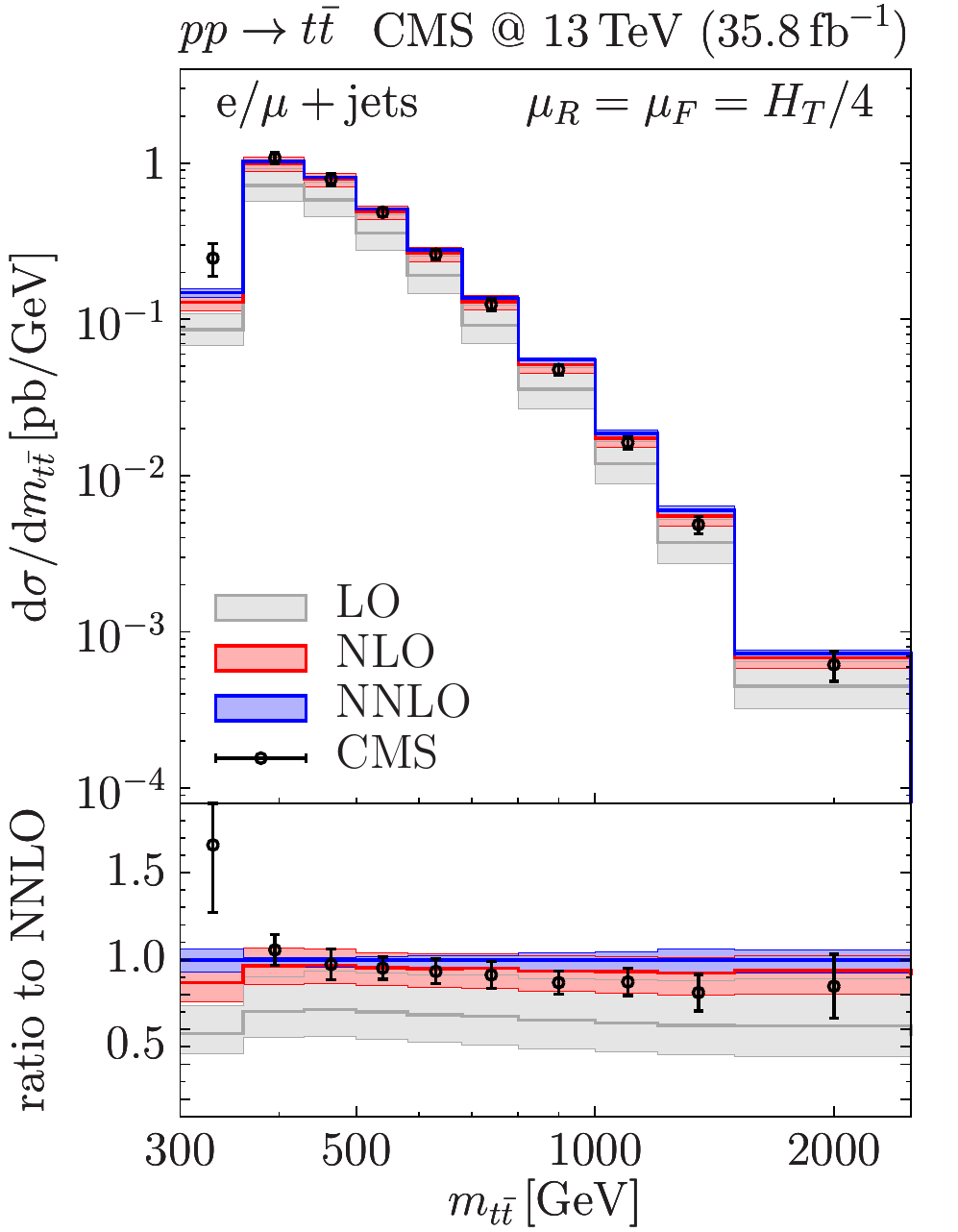}
\vspace{-4ex}
\caption{\label{fig:mtt}
  Single-differential cross sections as a function of $\mtt$.
  CMS data~\cite{Sirunyan:2018wem} and LO, NLO and NNLO results
  for central scales equal to $\Ht/2$ (left), $\mtt/2$ (central) and $\Ht/4$ (right).
}
\end{figure}

The invariant-mass distribution of the top-quark pair is reported in Fig.~\ref{fig:mtt}.
The distribution is peaked at \mbox{$\mtt \sim 400~{\rm GeV}$}.
The characteristic hard-scattering scale for this distribution is of the order of $\mtt$ itself.
We use our default scale choice
\mbox{$\mu_0=H_T/2$}~(left), and two other central values, namely \mbox{$\mu_0=\mtt/2$}~(central) and \mbox{$\mu_0=H_T/4$}~(right).

We first comment on the convergence of the perturbative series for the three scale choices.
In the cases \mbox{$\mu_0=H_T/2$} and \mbox{$\mu_0=\mtt/2$} we see that LO and NLO bands
do not overlap, analogously to what was previously observed for the $\pth$ distribution and for the total cross section in Eq.~(\ref{eq:sigmatot}).
In the case \mbox{$\mu_0=H_T/2$}, the NNLO corrections enhance the NLO result
by about $10\%$ in the peak region,
and their effect slightly increases with $\mtt$ up to about $15\%$ in the highest-$\mtt$ bin.
Using \mbox{$\mu_0=\mtt/2$}, the NNLO effect is similar in the peak region,
but it increases to about $20\%$ at high $\mtt$. The NLO and NNLO bands do overlap using both
\mbox{$\mu_0=H_T/2$} and \mbox{$\mu_0=\mtt/2$}.
As observed in Ref.~\cite{Czakon:2016dgf}, the choice \mbox{$\mu_0=H_T/4$} leads to a faster convergence
of the perturbative series.
However, in the region where $\mtt > 360~{\rm GeV}$ we see that the size of the scale-variation band is very much reduced in going from the NLO to the NNLO result.
This behaviour suggests that the central scale $\mu_0=H_T/4$ is (accidentally) quite close to a region of (local) minimal sensitivity \cite{Stevenson:1981vj} of the scale dependence of the NNLO result. In view of this feature, we think that the NNLO scale variation band
with $\mu_0=H_T/4$ likely underestimates the perturbative uncertainty in this $\mtt$ region and, especially, in the intermediate mass range $400~{\rm GeV}\ltap\mtt \ltap 1~{\rm TeV}$.

We now comment on the comparison with the data. The first bin, \mbox{$300\,{\rm GeV}<\mtt< 360\,{\rm GeV}$},
deserves a separate discussion.
The experimental result in this bin is significantly above the theoretical predictions,
independently of the scale choice.
This disagreement may have various origins.
We first note that the result of our calculation for on-shell top quarks is non-vanishing only in the limited region where
\mbox{$2\mt=346.6~{\rm GeV} <\mtt< 360~{\rm GeV}$}.
In this region the NLO and NNLO effects are largest, regardless of the scale choice.
Moreover, this region is particularly sensitive to the value of the top-quark mass.
If the top-quark mass is smaller\footnote{Incidentally, we note that in Ref.~\cite{Sirunyan:2019zvx}
the CMS collaboration uses NLO QCD results obtained with NNPDF31 PDFs to extract the value
$\mt=170.81 \pm 0.68$~GeV from a fit of a measurement of the triple-differential cross section
as a funtion of $\mtt$, $\ytt$ and the multiplicity of additional jets.}
by a couple of {\rm GeV}, the predicted cross section
in this bin becomes larger, without significantly affecting the NNLO result in each of the higher-$\mtt$ bins.
Another possible reason for the discrepancy may be the unfolding procedure~\cite{Sirunyan:2018wem}
that is used to convert the data from particle to parton level.
We expect such unfolding procedure to be particularly delicate in the threshold region:
both the actual value of the top-quark mass used in the extrapolation
and off-shell effects may have a significant impact.
Further perturbative QCD effects beyond NNLO may also be relevant,
but they are not expected to be large enough to explain the observed discrepancy with respect to the data.
Overall, excluding the first bin, the NNLO result for the central scale \mbox{$\mu_0=\mtt/2$} leads to
the best agreement with the data. The result for \mbox{$\mu_0=H_T/2$} is also consistent
with the data within uncertainties.

\begin{figure}
\includegraphics[height=0.30\textheight]{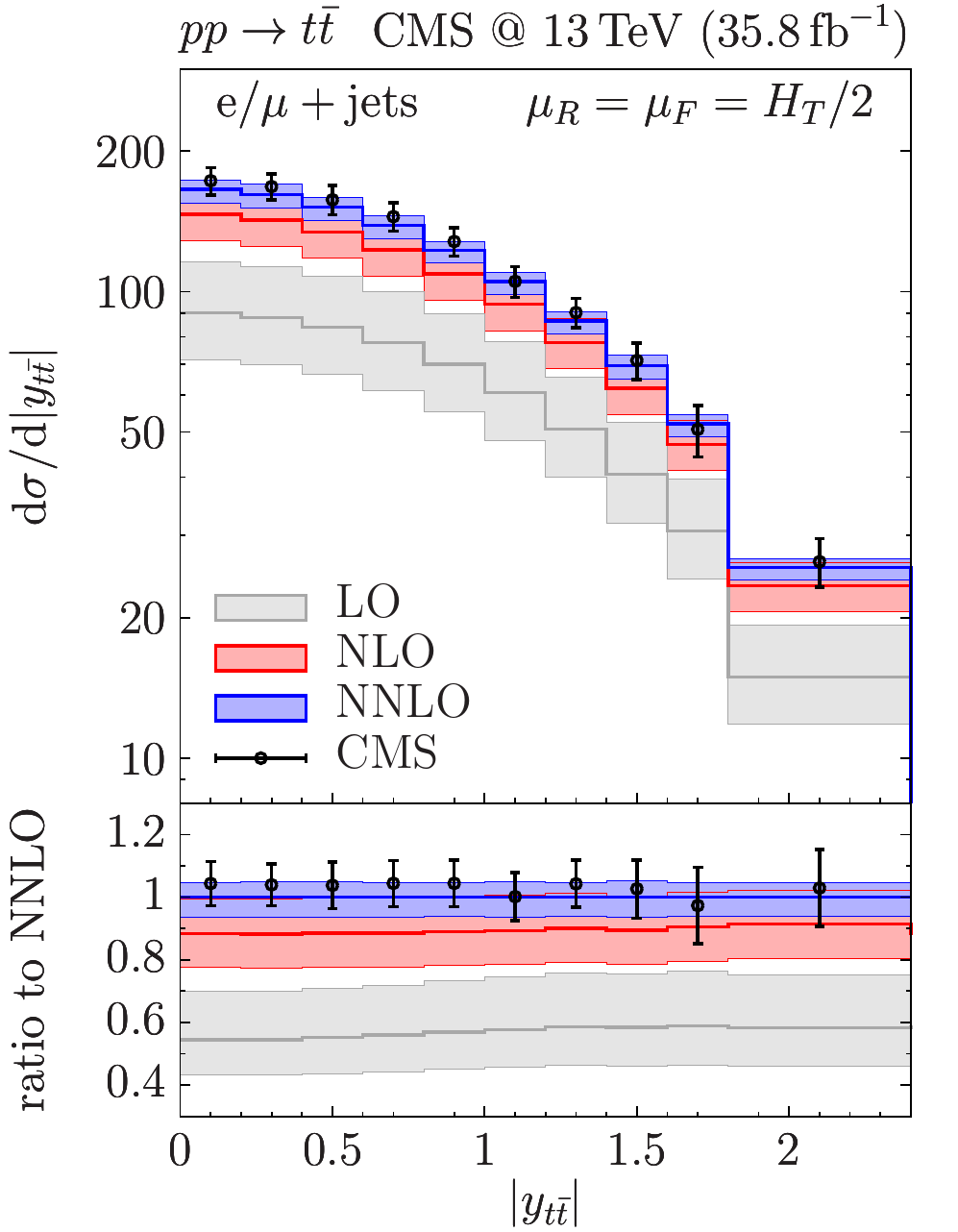}
\includegraphics[height=0.30\textheight]{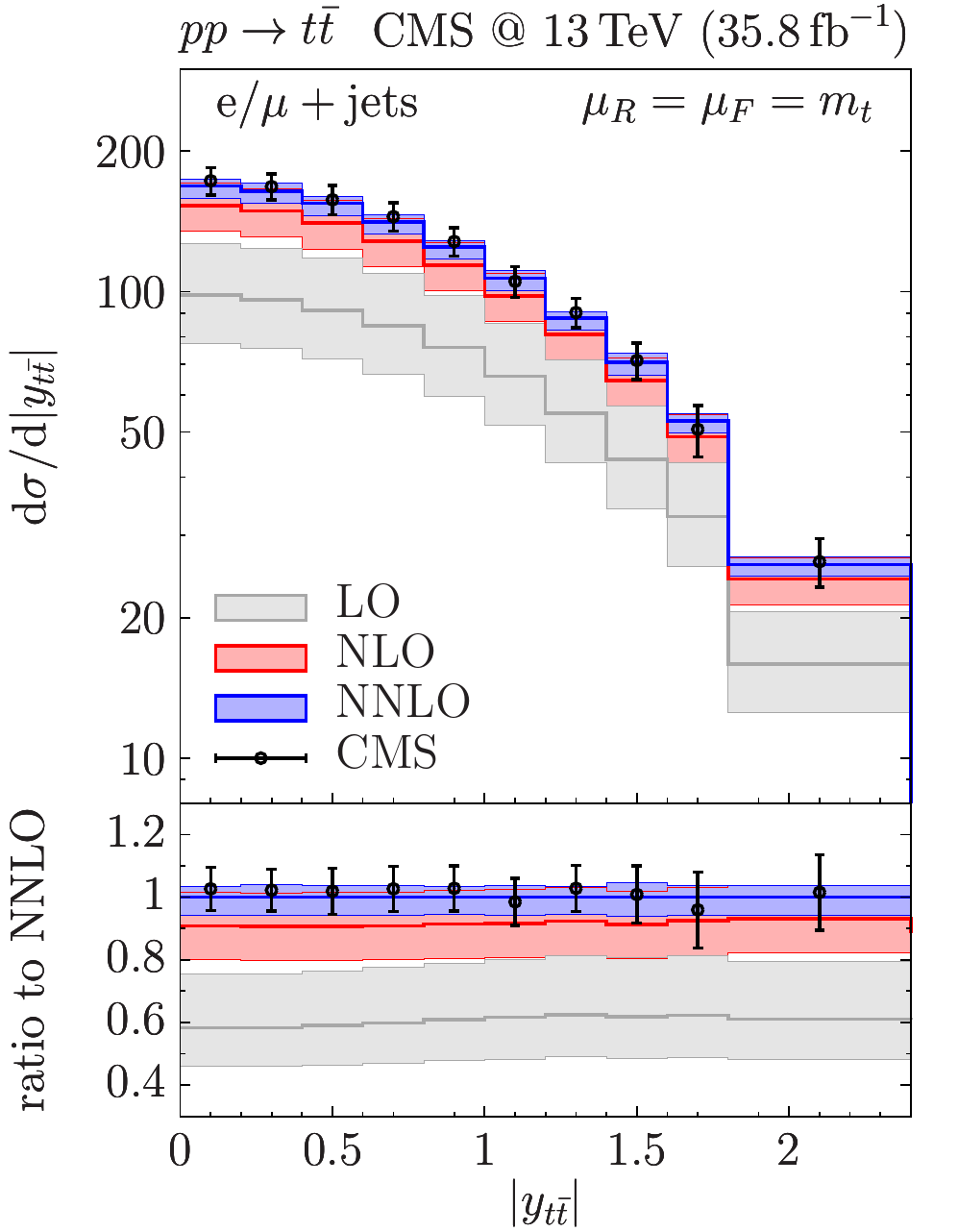}
\includegraphics[height=0.30\textheight]{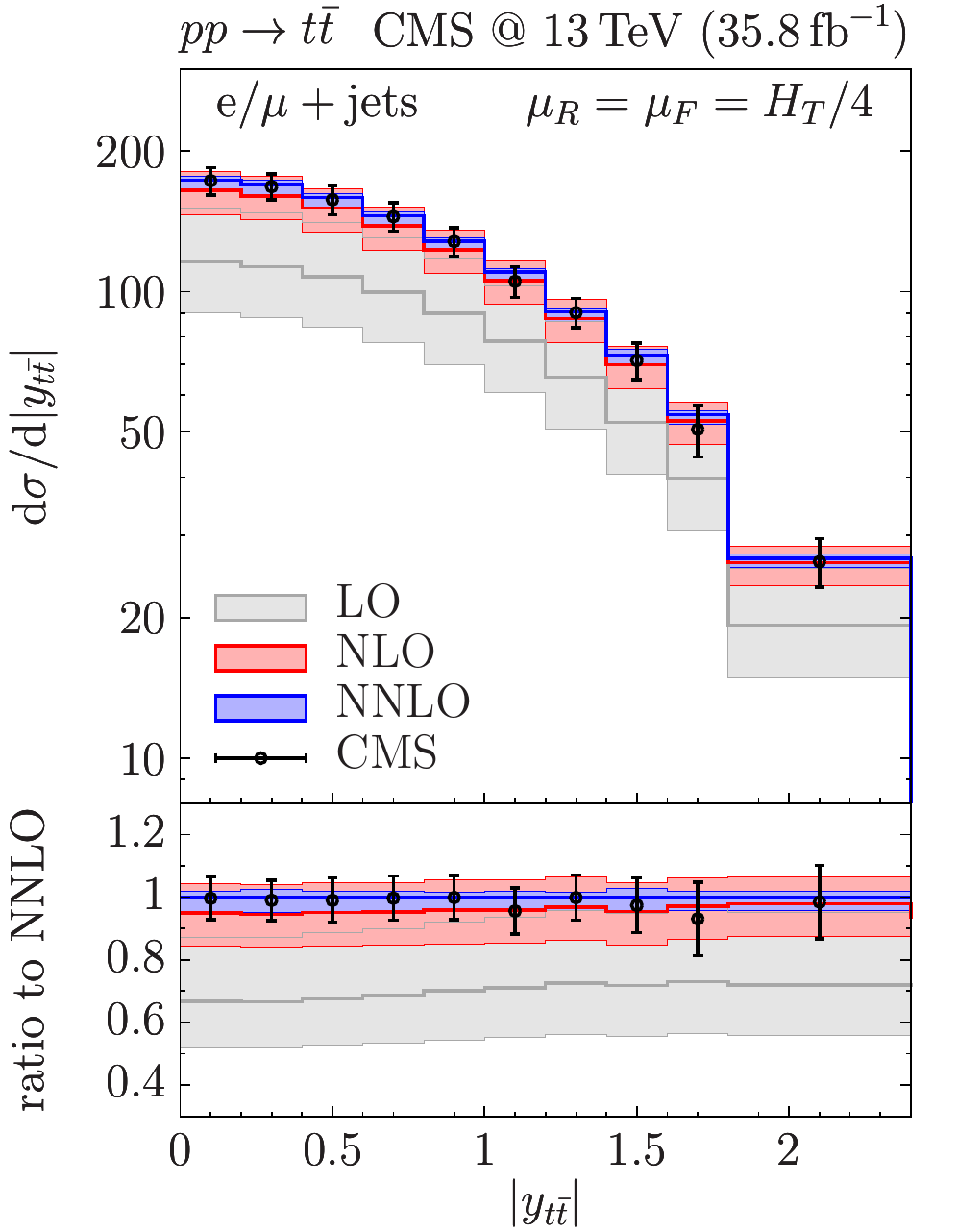}
\vspace{-4ex}
\caption{\label{fig:ytt}
  Single-differential cross sections as a function of  $\ytt$.
  CMS data~\cite{Sirunyan:2018wem} and LO, NLO and NNLO results
  for central scales equal to $\Ht/2$ (left), $\mt$ (central) and $\Ht/4$ (right).
}
\end{figure}

We finally consider the distribution in the rapidity of the top-quark pair, $\ytt$.
A natural scale choice for this distribution is the top-quark mass,
as in the case of the computation of the total cross section.
In Fig.~\ref{fig:ytt} we report the rapidity distribution of the $\ttb$ system for our default choice
\mbox{$\mu_0=H_T/2$} (left), for \mbox{$\mu_0=\mt$} (central) and for \mbox{$\mu_0=H_T/4$} (right).
We first observe that the choices \mbox{$\mu_0=H_T/2$} and \mbox{$\mu_0=\mt$} lead to rather similar results.
The LO and NLO bands marginally overlap, and the impact of NNLO corrections in the central rapidity region
is about $10\%$, consistently with what we find for the total cross section.
As previously observed for the $\mtt$ distribution, the choice \mbox{$\mu_0=H_T/4$} leads to a faster convergence
of the perturbative expansion \cite{Czakon:2016dgf},
and to a strong reduction of scale uncertainties at NNLO.
The data nicely agree with the NNLO predictions for all the three scale choices.

We conclude this section with a few comments on the effects of EW corrections
on the distributions considered so far.
The effect of the complete EW corrections to top-quark pair production was studied in
Ref.~\cite{Pagani:2016caq}, and NNLO QCD+NLO EW predictions for some differential observables
were presented in Ref.~\cite{Czakon:2017wor}.
In the case of the $\pth$ distribution, the EW corrections are negative and lead
to percent-level effects at low transverse momenta, which increase up to about $-5\%$
in the highest-$p_T$ bins in Fig.~\ref{fig:pth}.
Such effects are of the order of (or smaller than)
the residual perturbative uncertainties of the NNLO result,
and significantly smaller than the current experimental uncertainties.
An analogous conclusion can be drawn for the invariant-mass distribution,
which receives EW corrections of about $+2\%$ in the small-$\mtt$ region
and $-2\%$ in the highest-$\mtt$ region in Fig.~\ref{fig:mtt}.
The impact of EW corrections on the rapidity distributions is typically of ${\cal O}(1\%)$ or smaller.
Therefore, we conclude that the impact of EW corrections is not expected to significantly affect
our comparison with the data. Similar conclusions can be drawn for the double-differential distributions
that we present in the following.
More precise data, with higher reach in transverse momenta, will definitely call for the inclusion
of EW corrections~\cite{Czakon:2017wor}.

\subsection{Double-differential distributions}
\label{sec:double}

\begin{figure}[t]
\begin{center}
\includegraphics[height=0.30\textheight]{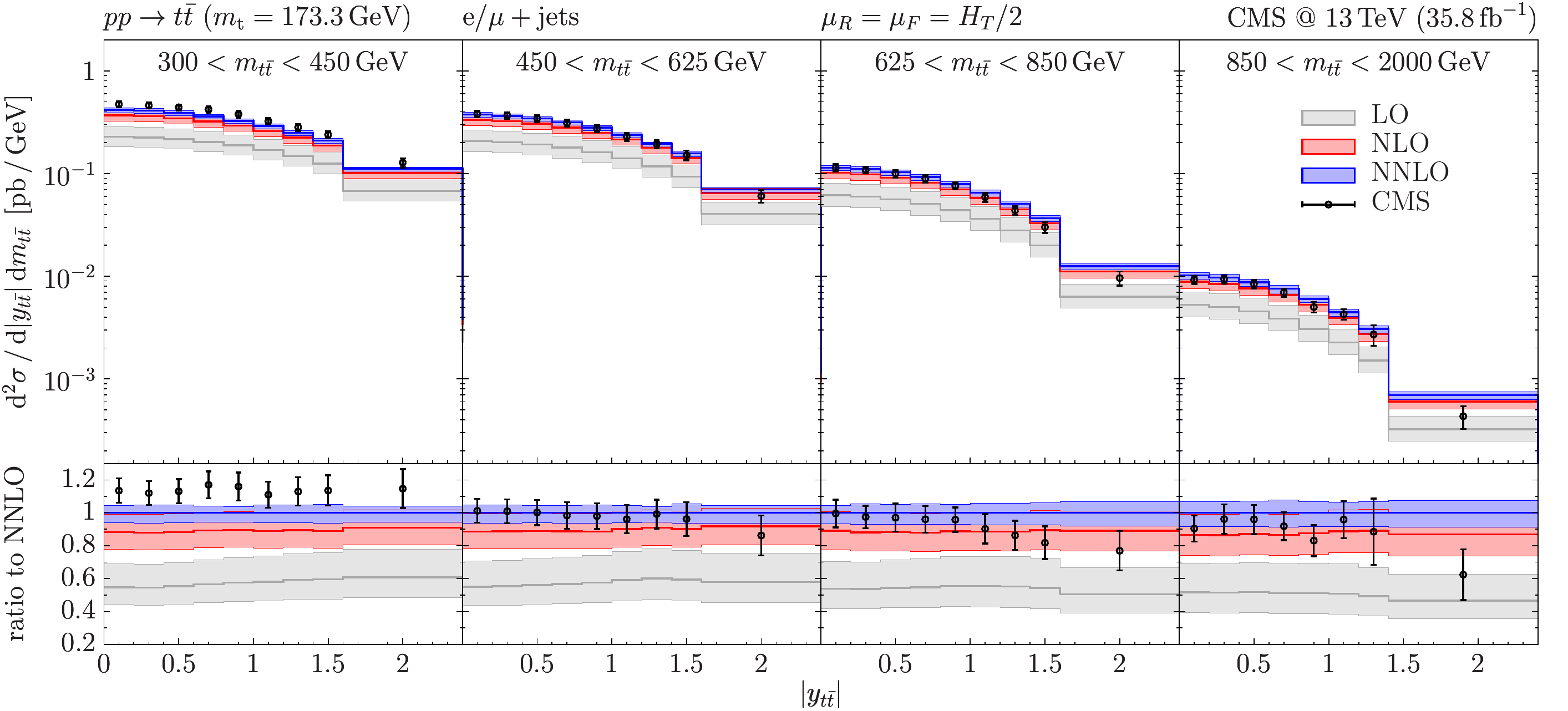}\\
\includegraphics[height=0.30\textheight]{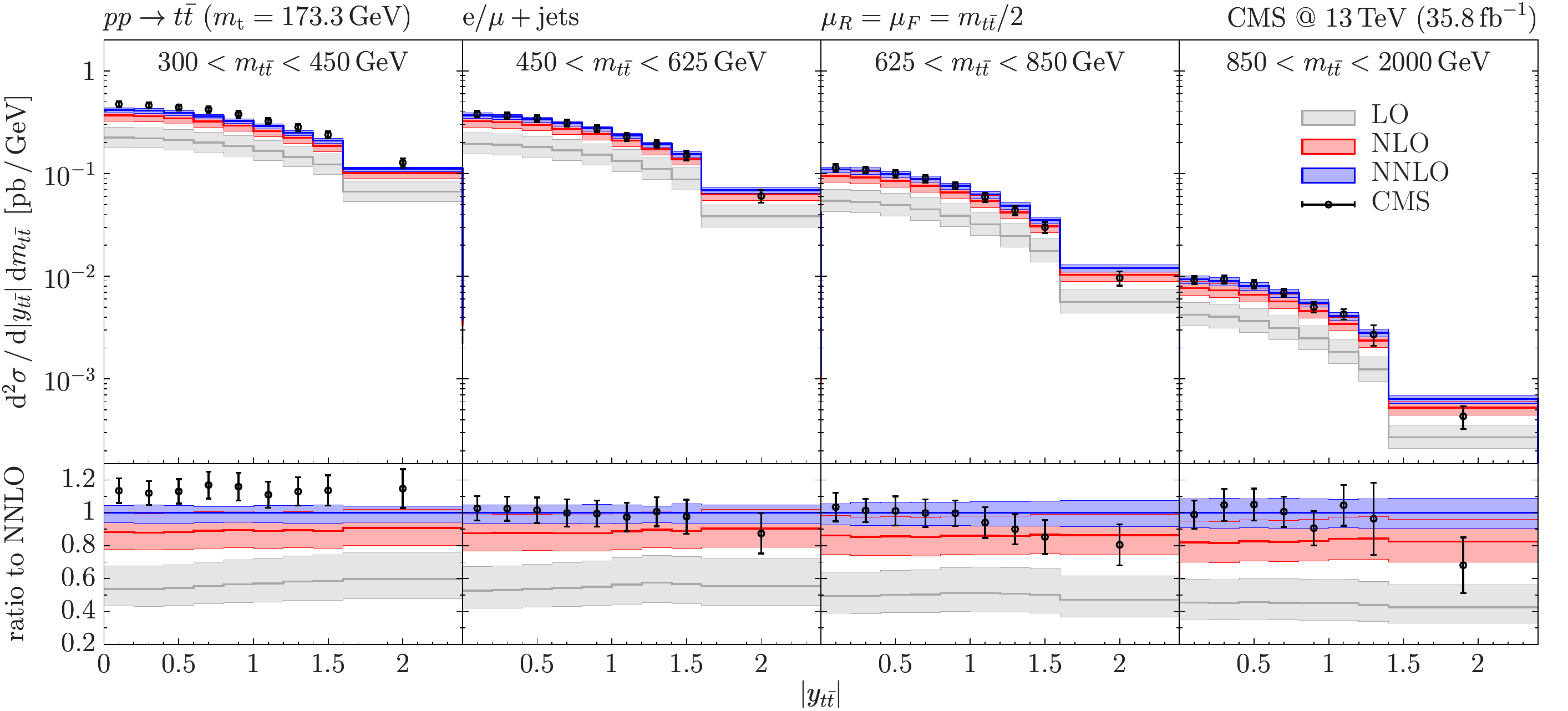}
\end{center}
\vspace{-4ex}
\caption{\label{fig:ytt_mtt}
  Double-differential cross sections as a function of $\ytt$ in four $\mtt$ intervals.
  CMS data~\cite{Sirunyan:2018wem} and LO, NLO and NNLO results
  for central scales $\Ht/2$ (upper) and $\mtt/2$ (lower).
}
\end{figure}

\begin{figure}[t]
\begin{center}
\includegraphics[height=0.30\textheight]{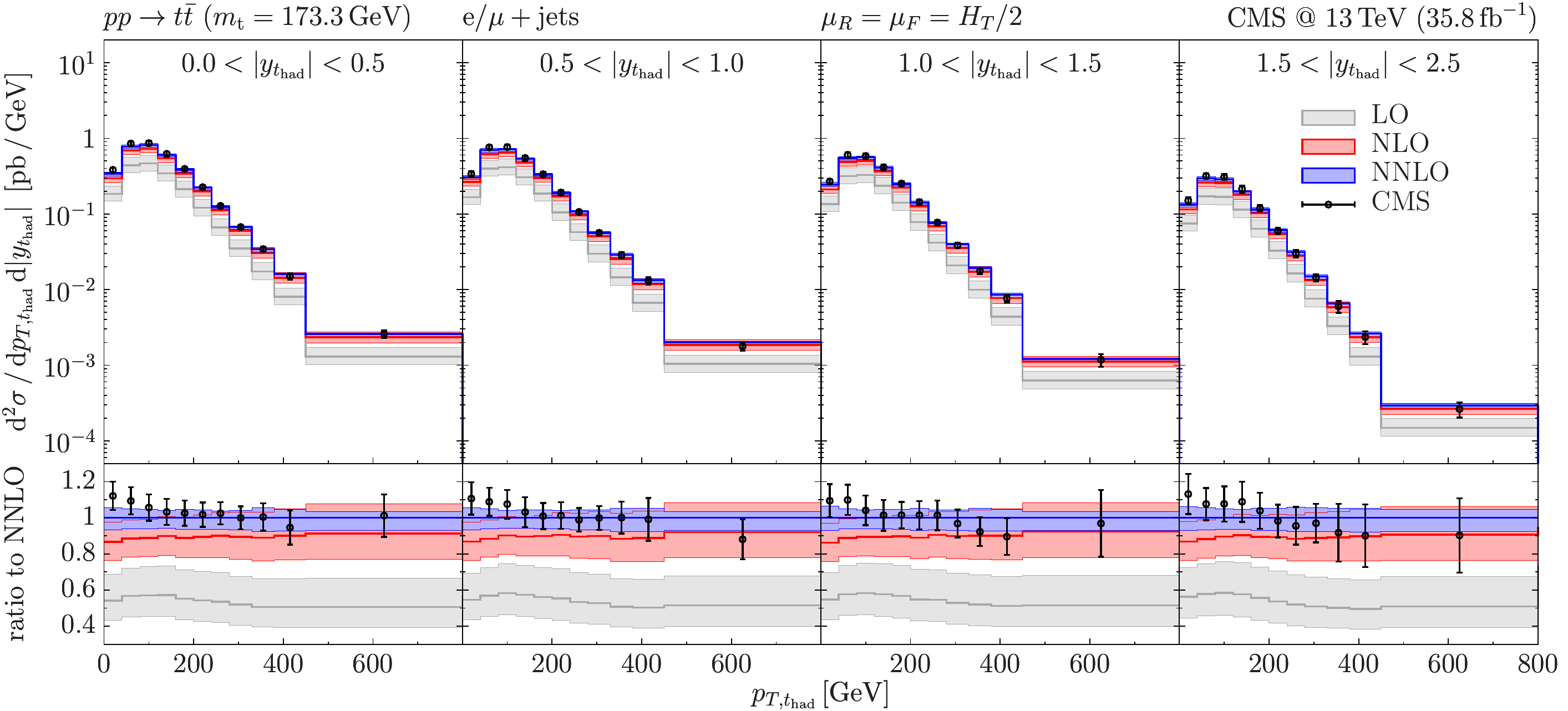}\\
\includegraphics[height=0.30\textheight]{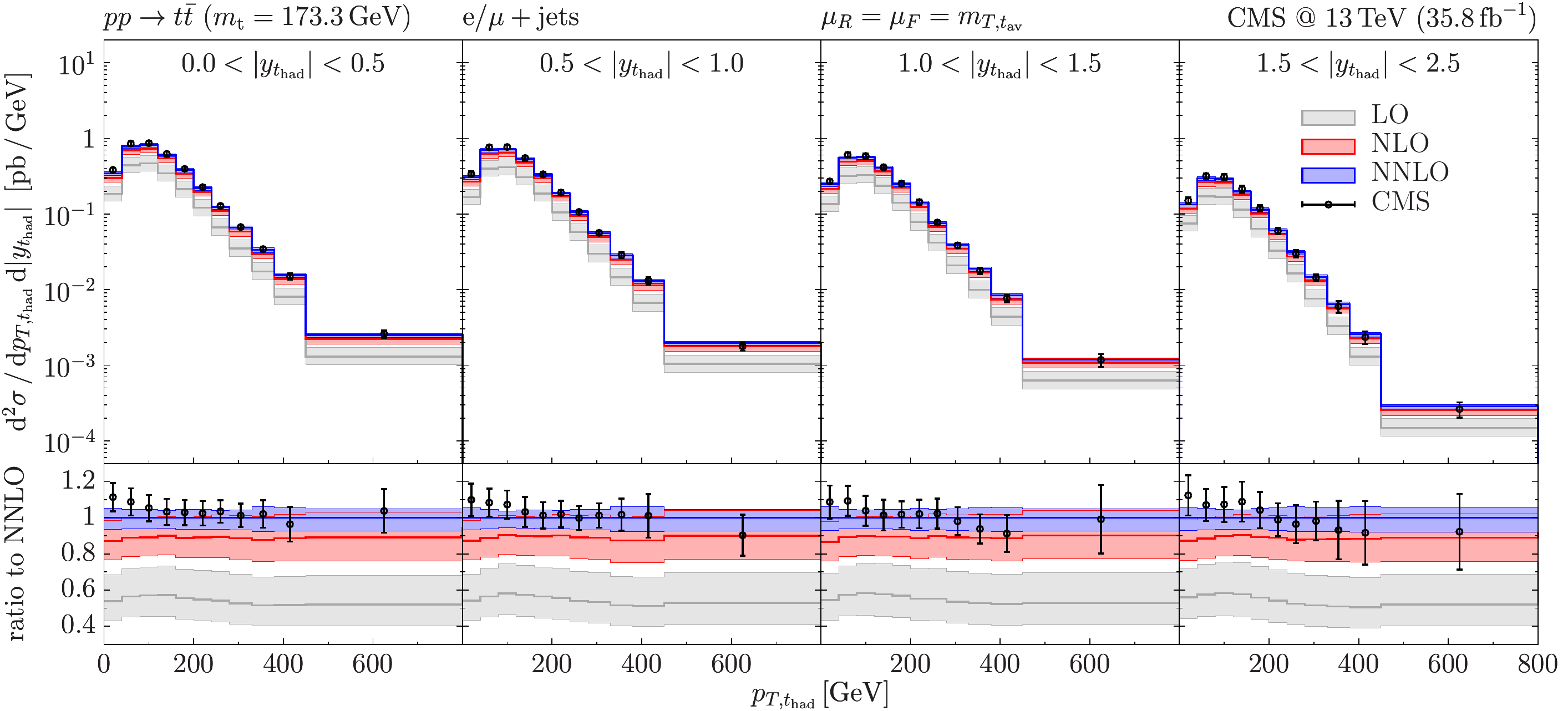}
\end{center}
\vspace{-4ex}
\caption{\label{fig:pth_yth}
  Double-differential cross sections as a function of $\pth$ in four $\yth$ intervals.
  CMS data~\cite{Sirunyan:2018wem} and LO, NLO and NNLO results
  for central scales $\Ht/2$ (upper) and $\mTav$ (lower).
}
\end{figure}

\begin{figure}[t]
\begin{center}
\includegraphics[height=0.30\textheight]{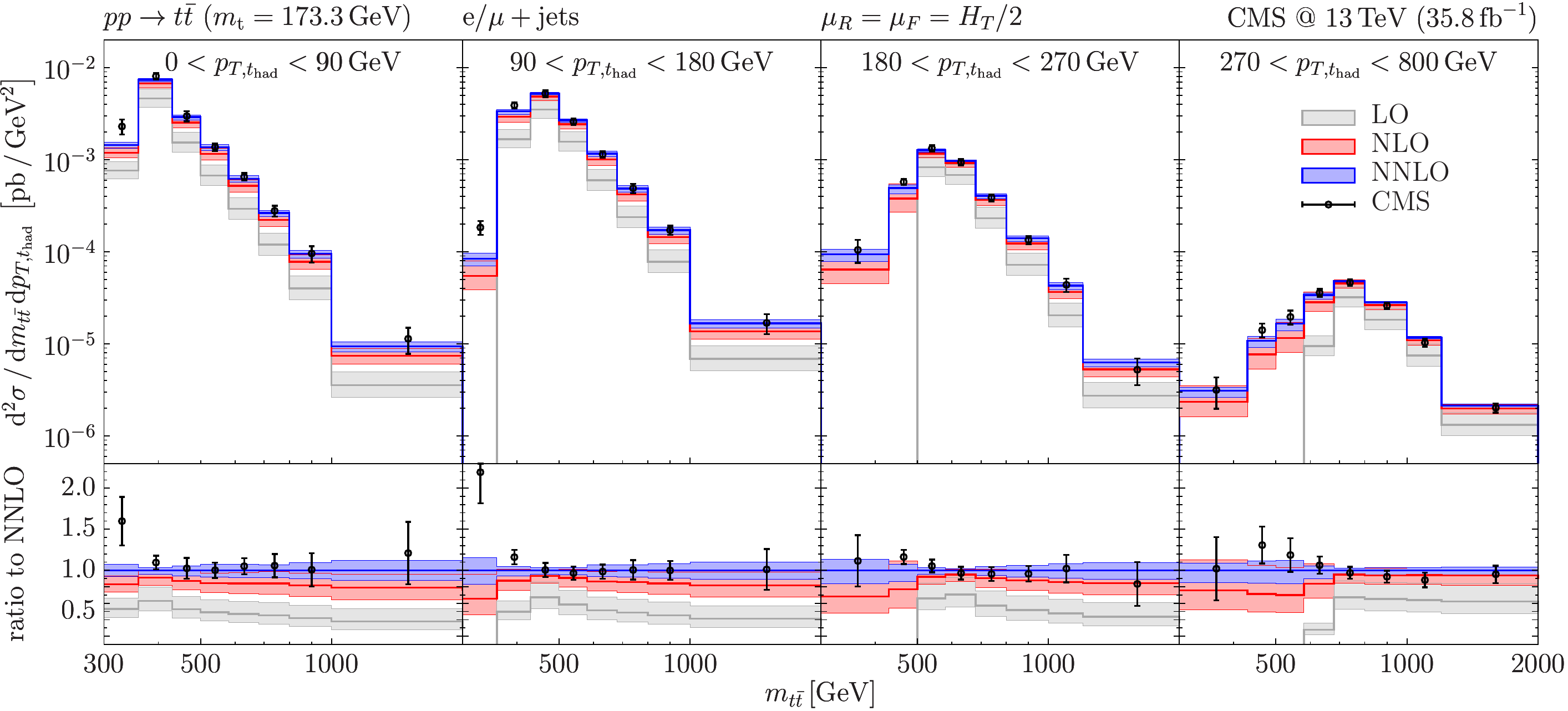}\\
\includegraphics[height=0.30\textheight]{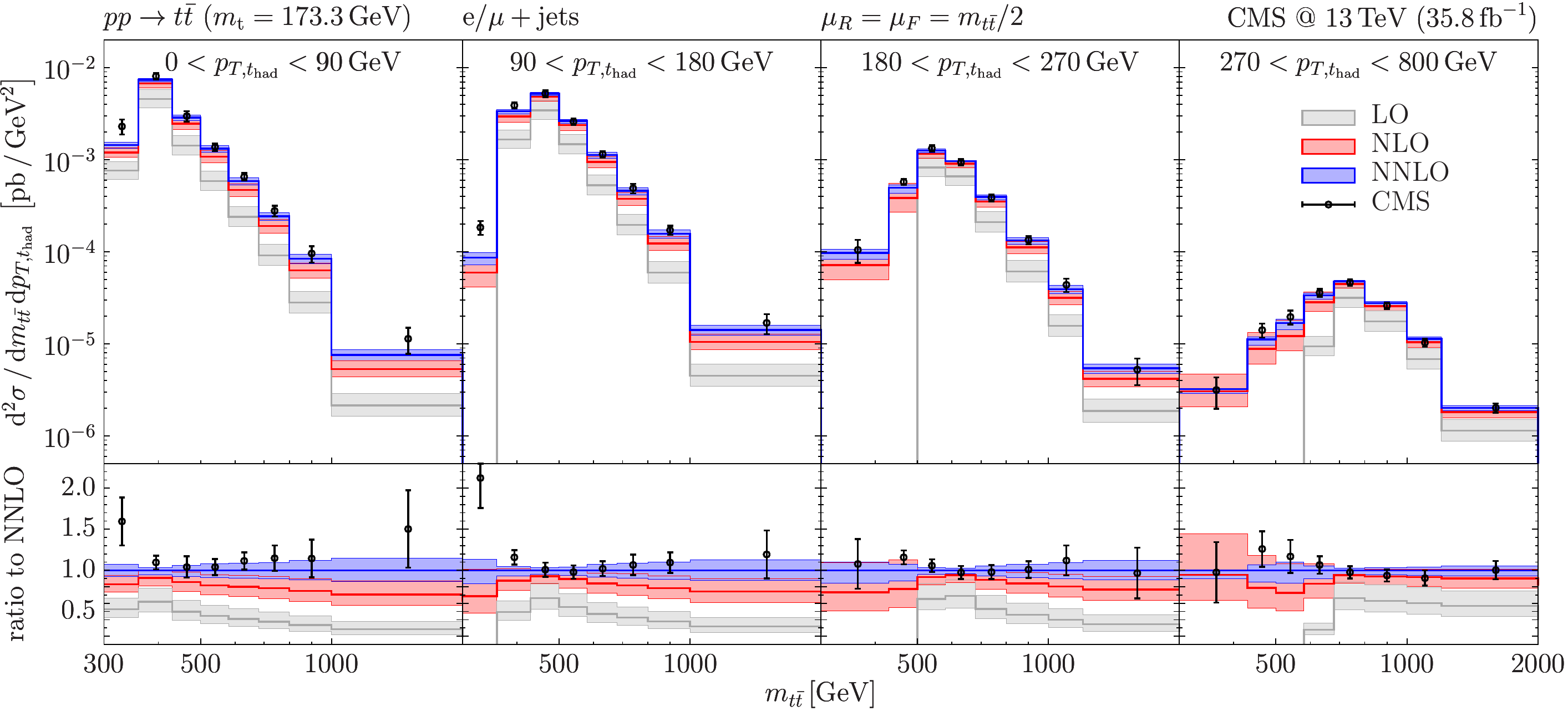}
\end{center}
\vspace{-4ex}
\caption{\label{fig:mtt_pth}
  Double-differential cross sections as a function of $\mtt$ in four $\pth$ intervals.
  CMS data~\cite{Sirunyan:2018wem} and LO, NLO and NNLO results
  for central scales $\Ht/2$ (upper) and $\mtt/2$ (lower).
}
\end{figure}

In this section we present results for double-differential distributions and compare them with the
CMS measurements from Ref.~\cite{Sirunyan:2018wem}.
We consider three double-differential distributions, namely
\begin{itemize}
\item the $\ytt$ distribution in $\mtt$ intervals~(see Fig.~\ref{fig:ytt_mtt});
\item the $\pth$ distribution in $\yth$ intervals~(see Fig.~\ref{fig:pth_yth});
\item the $\mtt$ distribution in $\pth$ intervals~(see Fig.~\ref{fig:mtt_pth}).
\end{itemize}
We present results for the default scale choice \mbox{$\mu_0=H_T/2$}
and 
at an additional scale of the order of the characteristic hard scale of the corresponding double-differential distribution.

The $\ytt$ distribution in four $\mtt$ intervals is presented in Fig.~\ref{fig:ytt_mtt}.
The central scales are \mbox{$\mu_0=H_T/2$}~(upper) and \mbox{$\mu_0=\mtt/2$}~(lower), and they lead to similar results, consistently with those for the single-differential cross sections in Figs.~\ref{fig:mtt} and \ref{fig:ytt}.
The impact of the radiative corrections is relatively uniform in $\ytt$ and in $\mtt$.
The scale \mbox{$\mu_0=\mtt/2$} leads to slightly larger radiative corrections in the highest-$\mtt$ interval.
In the first $\mtt$ interval, the NNLO prediction slightly undershoots the data,
consistently with what is observed in the low-$\mtt$ region in Fig.~\ref{fig:mtt}.
We note that, at variance with the first $\mtt$ bin in Fig.~\ref{fig:mtt},
here the first $\mtt$ interval extends up to $450$ GeV,
thereby leading to a better agreement between the NNLO prediction and the data.
At high values of $\mtt$, the scale \mbox{$\mu_0=\mtt/2$} leads to a slightly better agreement
with the data.
In the highest-$\ytt$ bin, for both central scales, the experimental result
is below the NNLO prediction in all invariant-mass intervals except the first one.

In Fig.~\ref{fig:pth_yth} we present the $\pth$ distribution in $\yth$ intervals,
where $\yth$ is the rapidity of the hadronically decaying top or antitop
quark with transverse momentum $\pth$.
The QCD results use the scales \mbox{$\mu_0=H_T/2$}~(upper) and \mbox{$\mu_0=\mTav$}~(lower) as a natural scale,
analogously to the $\pth$ distribution in Fig.~\ref{fig:pth}.
We observe that the two scale choices lead to very similar results.
The impact of the radiative corrections is rather uniform in the considered $\yth$ intervals.
Discussing the single-differential $\pth$ spectrum in Fig.~\ref{fig:pth},
we observed that the NNLO prediction is slightly harder than the data.
In Fig.~\ref{fig:pth_yth} we observe that this feature holds in all the $\yth$ intervals.
Nonetheless, the overall agreement between the NNLO prediction and the data is good.

We finally discuss the $\mtt$ distribution in $\pth$ intervals, which is shown in Fig.~\ref{fig:mtt_pth}.
We use the central scales \mbox{$\mu_0=H_T/2$}~(upper) and \mbox{$\mu_0=\mtt/2$}~(lower).
In each $\pth$ interval we have $\ptmin < \pth < \ptmax$.
The introduction of kinematical cuts on $\pth$ significantly affects the $\mtt$ distribution.
In the presence of a cut, $\ptmax$, on the maximum $p_T$ of the hadronically decaying top quark,
high values of $\mtt$ can be reached by either increasing the transverse momentum of the top quark
decaying to a leptonic final state, or increasing the rapidity separation \mbox{$|y_t - y_{\bar t}|$}.
These kinematical regions are both dynamically suppressed for heavy-quark production
and, therefore, $\ptmax$ produces a faster suppression at high invariant masses.
This effect can be observed by comparing the $\mtt$ single-differential distribution in Fig.~\ref{fig:mtt} with the present double-differential distribution.

The presence of a lower cut, $\ptmin$, on the minimum $\pth$ has an even more drastic effect on the $\mtt$ distribution,
as it imposes the constraint \mbox{$\mtt> 2\sqrt{\mtsquare+\ptmin^2} \equiv 2 \mtmin$} for LO kinematics.
Below this unphysical threshold the LO result vanishes,
and the NLO and NNLO results are effectively LO and NLO predictions, respectively.
As a consequence, they suffer from larger theoretical uncertainties,
which is reflected by the stronger scale dependence.
Above this threshold, the LO distribution sharply increases up to a kinematical peak
close to $2\mtmin$.
Owing to this LO behaviour, soft-collinear radiation produces shape instabilities~\cite{Catani:1997xc}
in this $\mtt$ region at each subsequent perturbative order.
The qualitative behaviour of these shape instabilities is completely analogous to that observed and discussed in Ref.~\cite{Catani:2018krb} (see Figs.~10 and 20 and related comments therein) in the case of diphoton production
in the presence of $p_T$ cuts.
These perturbative instabilities are localized in a very narrow region
around the LO threshold, and therefore their effect is smeared if a sufficiently large bin size is considered, as is the case for the differential distribution in Fig.~\ref{fig:mtt_pth}.

The comparison with the data shows that in the first two $\pth$ intervals
the NNLO prediction undershoots the data in the first $\mtt$ bin.
This discrepancy at low-$\mtt$ is not resolved in the two highest-$\pth$ intervals,
since the larger bin size ($300\text{ GeV} < \mtt < 430\text{ GeV}$) renders the distribution less sensitive to the behaviour close to the physical threshold.
These observations are consistent with the expectations
from the behaviour of the single-differential distribution in Fig.~\ref{fig:mtt} at low $\mtt$.
Excluding the narrower bins at low $\mtt$, the NNLO prediction in Fig.~\ref{fig:mtt_pth} is in very good agreement with the experimental measurements.
The results with \mbox{$\mu_0=H_T/2$}~(upper) and \mbox{$\mu_0=\mtt/2$}~(lower)
turn out to be rather similar, consistently with
our general expectation at the beginning of Section~\ref{sec:resu}.
The exception is the highest-$\pth$ interval, where the scale \mbox{$\mu_0=\mtt/2$} leads
to a quite large NLO scale dependence at low $\mtt$, which is drastically reduced at NNLO.
In this region of low $\mtt$ and high $\pth$, the invariant mass $\mtt$ is not the characteristic hard scale anymore:
the scale choice \mbox{$\mu_0=\mtt/2$} is not expected to be optimal,
and the scale \mbox{$\mu_0=H_T/2$} turns out to be more appropriate.

\section{Summary and outlook}
\label{sec:summary}

In this paper we have presented a new fully differential NNLO calculation
of top-quark pair production at hadron colliders.
The calculation is carried out by using the $q_T$ subtraction formalism
to handle IR divergences from real and virtual contributions, and it is implemented in the \Matrix{} framework.
Our code enables fast and efficient calculations of fiducial cross sections
and multi-differential distributions.

We have computed several single- and double-differential distributions of the top quarks,
and we have compared our results with recent measurements performed by the CMS collaboration
in the lepton+jets decay channel.
We have considered several values of the renormalization and factorization scales to compute each of the distributions.
We have used natural scales (i.e. $\mt$, $\mtt/2$ and the relevant transverse masses $m_T$)
of the order of the characteristic hard scale of the computed distribution,
and we have shown that the corresponding results are similar to what is obtained 
with the overall choice $\mu_0 = H_T/2$.
We find that both the natural scales and $\mu_0 = H_T/2$ lead
to a reasonable perturbative behaviour for all the distributions that we have examined.
The NNLO corrections substantially reduce the uncertainties of the theoretical predictions,
and they improve the overall agreement with the experimental measurements.
The largest deviation between data and the NNLO result
occurs close to the $\mtt$ threshold in single- and double-differential distributions.
This discrepancy could be related to a variety of effects,
including issues in the extrapolation of the data from particle to parton level,
which is expected to be delicate in such threshold region.
A lower value of the top-quark mass also has a significant impact close to the threshold.

The code that is used to perform these calculations is going to become public in a future \Matrix{} release,
providing a fast and flexible tool to compute (multi-)differential distributions
with arbitrary cuts on the top-quark kinematical variables.
The inclusion of NLO EW corrections and of top-quark decays is left to future work.

\vspace*{2ex}
\noindent {\bf Acknowledgements}

\noindent We are very grateful to Hayk Sargsyan for his contribution at early stages of this work.
We are also indebted to Federico Buccioni, Jean-Nicolas Lang, Jonas Lindert and Stefano Pozzorini for their ongoing support with {\sc OpenLoops~2}.
We wish to thank Ben Kilminster and Florencia Canelli for useful discussions.
This work is supported in part by the Swiss National Science Foundation (SNF) under contract 200020-169041. The work of SK is supported by the ERC Starting Grant 714788 REINVENT.

\newpage

\appendix

\section{Appendix}
\label{sec:validation}

The calculation presented in this paper represents the first complete application
of the $q_T$ subtraction formalism to the NNLO computation of (multi-)differential cross sections for the production of a colourful final state.
As a validation of our results, in this appendix we carry out a detailed comparison
of our predictions with those available for $\ttb$ production at NNLO.

In Ref.~\cite{Bonciani:2015sha}, the $q_T$ subtraction results for the contributions of the flavour off-diagonal partonic channels to the $\ttb$ total cross section
were compared with those of the numerical program {\sc Top++}~\cite{Czakon:2011xx}
by considering $p{\bar p}$ collisions at $\sqrt{s}=2$~TeV and $pp$ collisions
at $\sqrt{s}=8$~TeV.
In Ref.~\cite{Catani:2019iny}, the comparison with {\sc Top++} results
at the LHC energies $\sqrt{s}=8$ and
13~TeV was performed by considering all the partonic channels, and we found agreement
within the numerical uncertainties.
We have successfully repeated this comparison in
$p{\bar p}$ collisions at Tevatron energies and in $pp$ collisions
with centre-of-mass energies $\sqrt{s}$ up to 100~TeV.
We thus conclude that our calculation of the total cross section is in perfect agreement with the calculation of
Refs.~\cite{Baernreuther:2012ws, Czakon:2012zr, Czakon:2012pz, Czakon:2013goa}.
The typical computing time needed to obtain ${\cal O}(0.1\%)$
precise inclusive cross sections (including scale uncertainties) with our program is about 1000 CPU days.

We point out that the only ingredient that is not computed independently in our implementation and in the calculation of
Refs.~\cite{Baernreuther:2012ws, Czakon:2012zr, Czakon:2012pz, Czakon:2013goa} is
the finite part of the two-loop virtual amplitudes \cite{Czakon:2008zk,Baernreuther:2013caa}.
This contribution, however, turns out to have a very small quantitative impact,
namely ${\cal O}(0.1\%)$ of the NNLO total cross section at \mbox{$\sqrt{s}=13$~TeV}.

In the following we present a comparison of our NNLO differential results with the analogous results
from Ref.~\cite{Czakon:2016dgf}.
To this purpose, we exactly follow the setup therein: we consider $pp$ collisions
at \mbox{$\sqrt{s}=13$~TeV}, and 
we use CT14NNLO~\cite{Dulat:2015mca} PDFs.
The QCD running of $\as$ is evaluated at three-loop order with \mbox{$\as(m_Z)=0.118$},
and the pole mass of the top quark is fixed to \mbox{$\mt=173.3$~GeV}.
We consider four differential distributions: invariant mass ($\mtt$)
and absolute rapidity ($|\ytt|$) of the top-quark pair, and averages of the transverse momenta ($\ptav$)
and of the absolute rapidities ($|\yav|$) of the top and the antitop quark.
\begin{figure}
\begin{center}
\includegraphics[width=.45\textwidth]{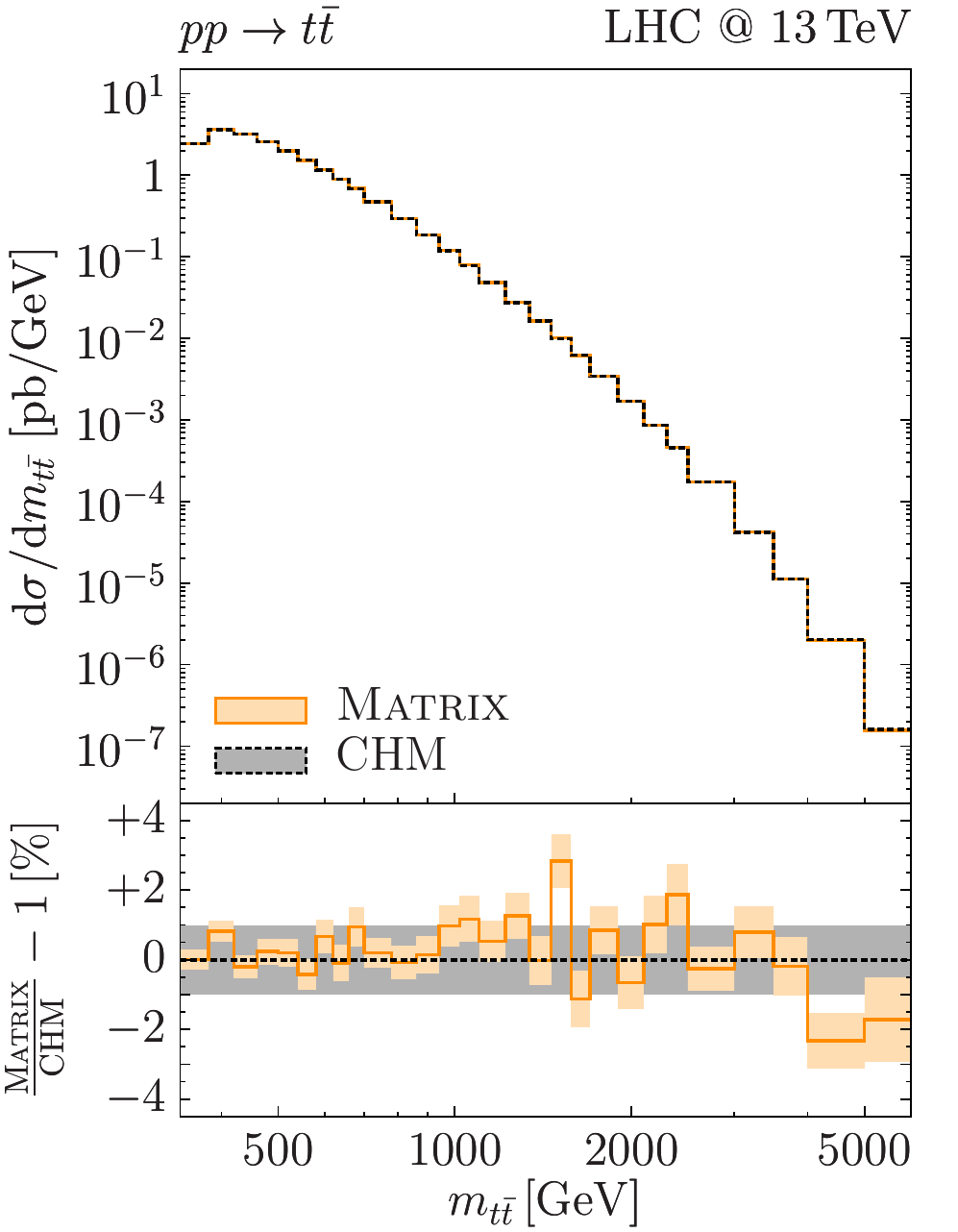}
\hspace*{0.5cm}
\includegraphics[width=.45\textwidth]{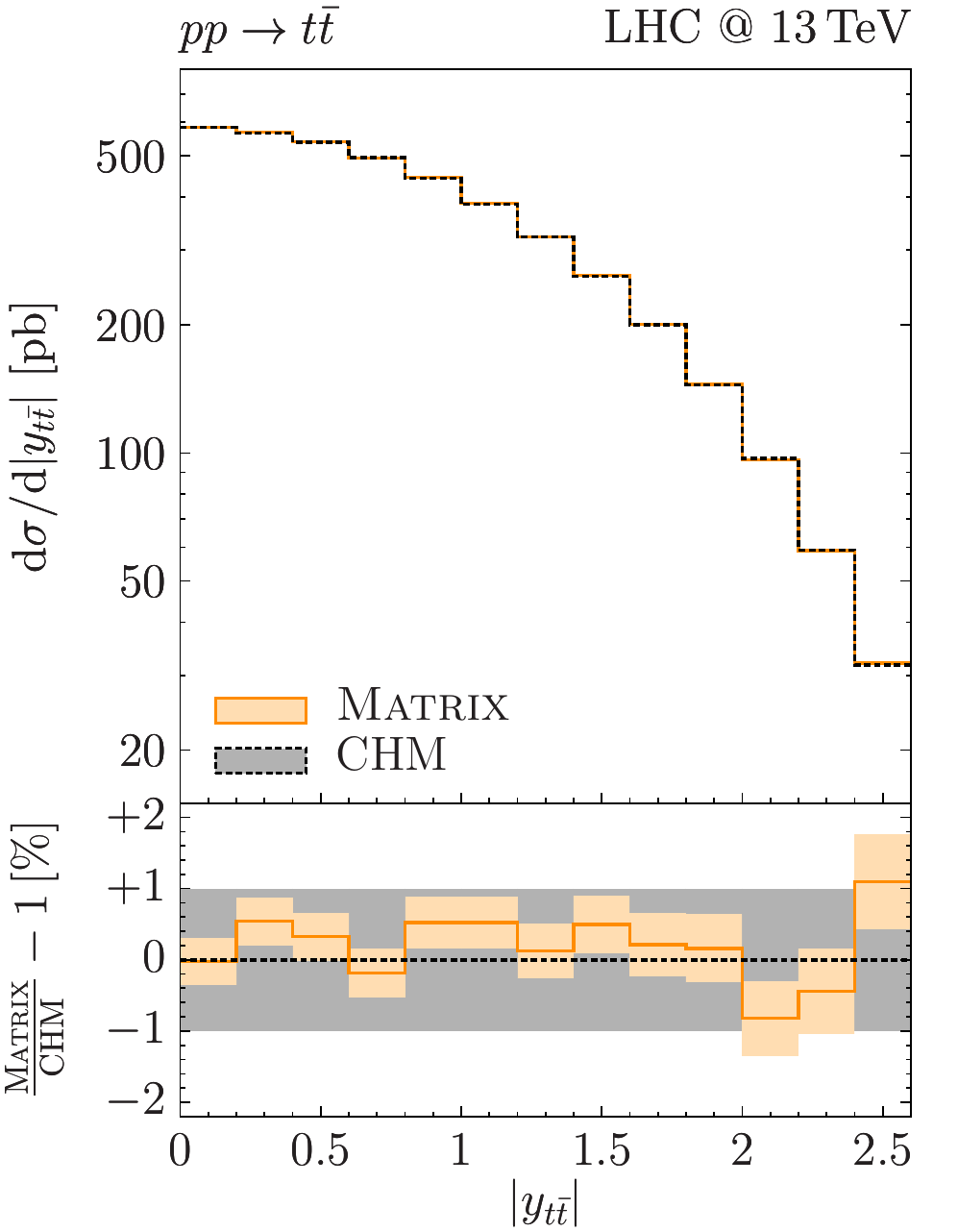}
\\
\vspace*{0.5cm}
\includegraphics[width=.45\textwidth]{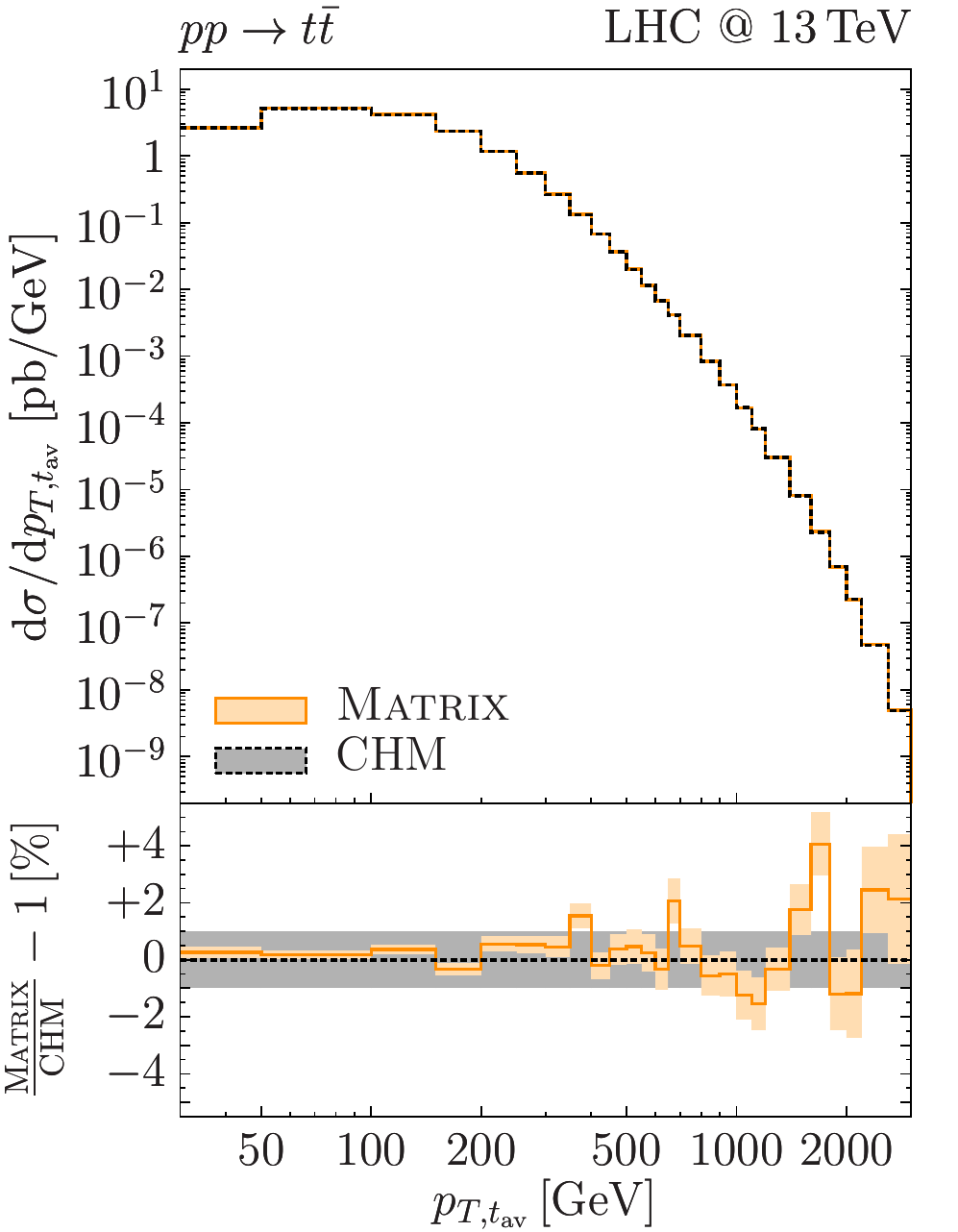}
\hspace*{0.5cm}
\includegraphics[width=.45\textwidth]{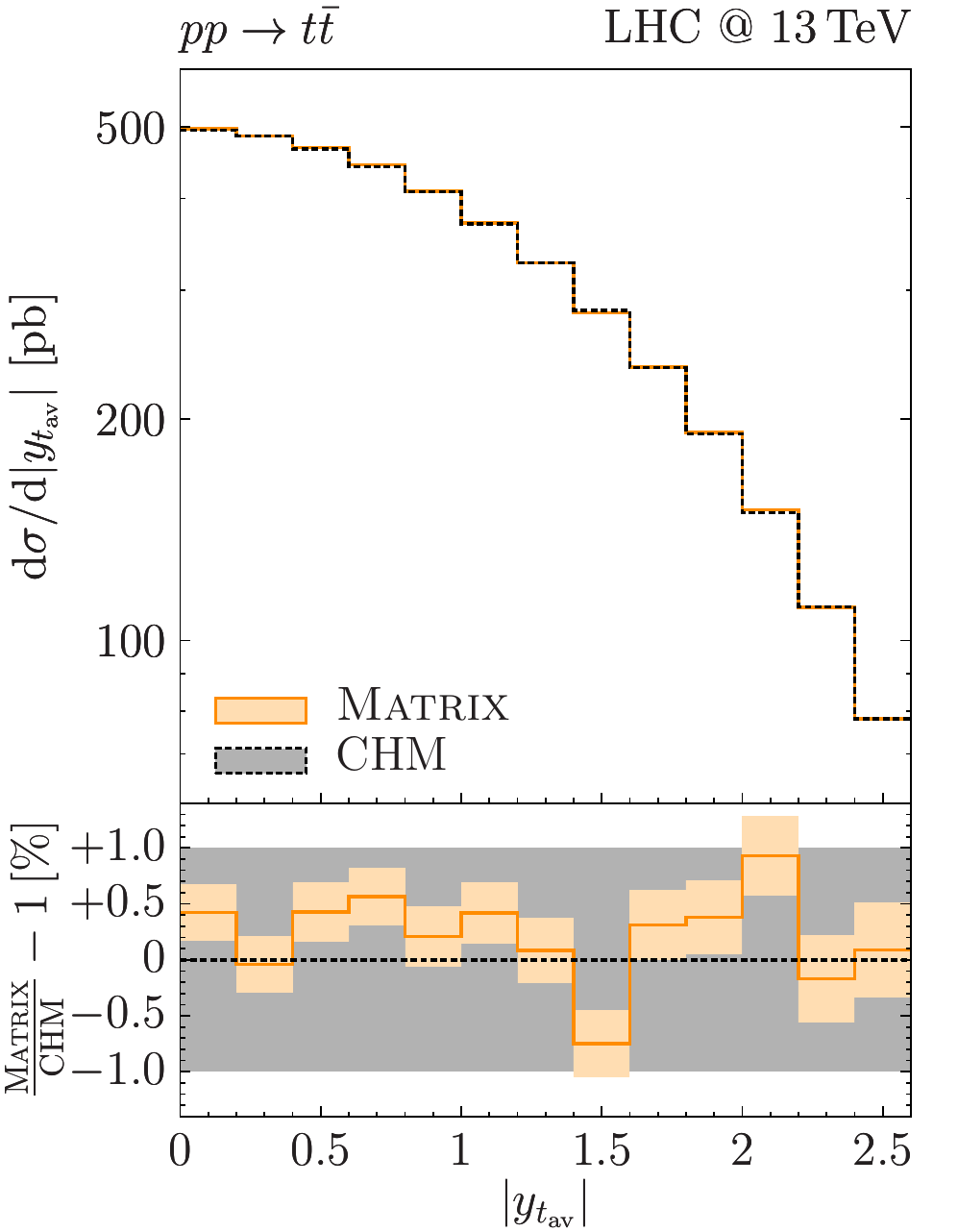}
\end{center}
\vspace{-2ex}
\caption{\label{fig:valid}
  Comparison between the NNLO differential distributions obtained with {\sc Matrix} (orange)
  and the results from Ref.~\cite{Czakon:2016dgf} (CHM, grey).
  The orange bands indicate the numerical uncertainty of our results, which is typically well below $1\%$.
  The grey band approximates the statistical uncertainty expected
  for the results of Ref.~\cite{Czakon:2016dgf}, based on corresponding comments in the text therein.
}
\end{figure}
The values of the renormalization and factorization scales are fixed to different values
for the different distributions.
In case of the $\mtt$, $|\ytt|$ and $|\yav|$ distributions the scales are set to $\Ht/4$.
In case of the $\ptav$ distribution, each of the $p_{T,t}$ and $p_{T,{\bar t}}$ distributions
is calculated with both scales set to half of the corresponding transverse mass.
This scale choice corresponds to \mbox{$\mu_0=\mTav/2$}
in the notation of Section~\ref{sec:single}.

The comparison of our results with those of Ref.~\cite{Czakon:2016dgf} is illustrated in Fig.~\ref{fig:valid}. 
In each case the upper panel shows the respective distribution, while the lower panel reports the ratio with respect
to the corresponding result from Ref.~\cite{Czakon:2016dgf}~(CHM).
Our results are stated with their numerical uncertainties.
The computing time needed to obtain our results is about 2000 CPU days,
mainly to achieve good statistical precision in the tails of the kinematical distributions.
The results from Ref.~\cite{Czakon:2016dgf} are quoted without an associated statistical uncertainty.
However, Ref.~\cite{Czakon:2016dgf} states that
``the narrowest bins possible'' are chosen in order to
``keep the Monte Carlo integration error within about 1\% in almost all bins''. 
Correspondingly, the approximate uncertainty of $\pm1\%$ is reported in the plots for reference.
We see that the agreement for all the considered distributions is excellent,
also in the extreme kinematical regions, i.e. large rapidities, large transverse momenta and
large invariant mass of the $\ttb$ pair.

\bibliography{biblio}

\end{document}